\documentclass{elsart}
\usepackage{amsfonts,amssymb,amsmath,amsbsy,amscd}
\usepackage{latexsym,euscript,exscale,epic,eepic,epsfig}
\usepackage{subfigure}
\usepackage{times}[9pt]
\newcommand{\eg}{e.g., }           % e.g.
\newcommand{\coll}{et al.}            % et al.
\begin{document}
\begin{frontmatter}
\title{On probabilistic aspects in the dynamic degradation of ductile materials}

\author[lr]{Herv\'e Trumel}
\footnote{Formerly at LMPM, ENSMA (UMR CNRS 6617),
T\'{e}l\'{e}port 2, BP 40109, 1 avenue Cl\'{e}ment Ader, F-86961
Futuroscope Chasseneuil Cedex. Corresponding author.},
 \ead{herve.trumel@cea.fr}
\author[lmt]{Fran\c cois Hild},
 \ead{hild@lmt.ens-cachan.fr}
\author[va]{Gilles Roy},
 \ead{gilles.roy@cea.fr}
\author[dif]{Yves-Patrick Pellegrini},
 \ead{yves-patrick.pellegrini@cea.fr}
\author[dif]{Christophe Denoual}
 \ead{christophe.denoual@cea.fr}

\address[lr]{CEA, DAM, Le Ripault, F-37260 Monts, France.}
\address[lmt]{LMT-Cachan, ENS Cachan/CNRS/UPMC/PRES UniverSud Paris, 61 avenue du Pr\'{e}sident Wilson, F-94235 Cachan Cedex, France.}
\address[va]{CEA, DAM, Valduc, F-21120 Is-sur-Tille, France.}
\address[dif]{CEA, DAM, DIF, F-91272 Arpajon, France.}


\begin{abstract}
Dynamic loadings produce high stress waves leading to the
spallation of ductile materials such as aluminum, copper,
magnesium or tantalum. The main mechanism used herein to explain
the change of the number of cavities with the stress rate is
nucleation inhibition, as induced by the growth of already
nucleated cavities. The dependence of the spall strength 
and critical time with the loading rate is investigated in 
the framework of a probabilistic model. The present approach, 
which explains previous experimental findings on the strain-rate dependence of the spall strength, is applied to analyze experimental data on tantalum.
\end{abstract}
\end{frontmatter}

%%%%%%%%%%%%%%%%%%%%%%%%%%%%%%%%%%%%%%%%%%%%%%%%%%%%%%%%%%%%%%%%%%%%%%%%%%%%%%%%%%%%%
%%%%%%%%%%%%%%%%%%%%%%%%%%%%%%%%%%%%%%%%%%%%%%%%%%%%%%%%%%%%%%%%%%%%%%%%%%%%%%%%%%%%%
%%
%% PAGE 1
%%
\section{Introduction}
\label{sec:Introduction}
The impact of a projectile on a target generates two shock waves
propagating in opposite directions. Meeting free surfaces, these
shock waves reflect back as two release waves, which generally
meet together at a definite location, the spall plane. Their
superposition produces a triaxial tensile ramp loading that often
results, prior to fracture, in  the nucleation, growth, and
coalescence of microvoids in most metals. This phenomenon is known
as ``ductile spalling.''

Although discovered long ago and studied by many authors (see
Meyers and Aimone, 1983; Curran \coll, 1987; Grady, 1988 for
reviews), its modeling still raises open questions. Since the
pioneering works of Carroll and Holt (1972) and of Glennie (1972),
void growth has by far been the main concern. This led many
authors to derive elastic-viscoplastic damage models using the
overall porosity as damage variable (see, \eg Eftis and Nemes,
1991; Cortes 1992), often comparable to the quasi-static class
of Gurson-like models (Gurson, 1977; Tvergaard, 1999). In these
models, nucleation and coalescence are generally dealt with in an
empirical fashion. In the recent years, however, renewed attention
has been paid to these processes. The present paper aims at
addressing the question of nucleation, coalescence being put aside
for future work (the interested reader may refer to some recent
works on this topic by Thomason, 1999; Tonks \coll, 2001;
Bontaz-Carion and Pellegrini, 2006).
\begin{figure}[t]
    \centering
    \includegraphics[width=10cm]{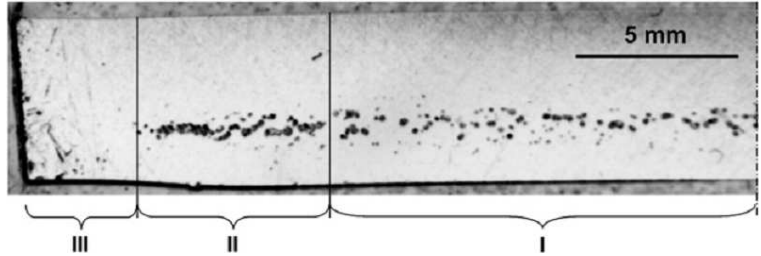}
    \caption{Example of a 5 mm thick tantalum sample damaged by a
    symmetric impact at $270$~m/s by a CuC2 flyer plate (the shock wave traveled from top to bottom); zone I: uniaxially loaded, zone II: biaxially loaded, zone III: rapidly unloaded. Only the left half of the target is shown. The right edge of the image is close to the symmetry axis.}
    \label{fig:Figure1}
\end{figure}

Some recent interrogations in relation to the definition of a
dynamic representative volume element (Roy, 2003; Dragon and
Trumel, 2003) seem to indicate that the overall porosity is not a
sufficient parameter, and that the entire void size
%%
%% PAGE 2
%%
distribution should be accounted for. The question of micro-inertia,
neglected for a long time, is the subject of a continued effort
(Ortiz and Molinari, 1992; Tong and Ravichandran, 1995;
Wang and Jiang, 1997; Molinari and Mercier, 2001;
Wu \coll, 2003; Roy, 2003). Not only does it slow down the growth
of individual voids, but also does it confine each void within
an evolving neighborhood bounded by an elastic relaxation wave.
Hence, dynamic void interactions are strongly linked to intervoid
spacing, itself driven by the nucleation process. The latter
thus appears as a crucial mechanism. This is
all the more the case that Roy (2003), studying pure tantalum over
a large range of shock levels and durations, showed extreme size
distributions to be present in recovered samples, indicating that
nucleation is a continuous process taking place up to coalescence.
Fig.~\ref{fig:Figure1} shows a tantalum sample recovered after
an impact at 270 m/s by a copper flyer plate (Roy, 2003), and
containing isolated voids up to about 100~$\mu$m in diameter
(even larger voids can be observed at lower impact velocities).
A detailed account of the nucleation process is clearly beyond the
present state of knowledge, although much progress is being made
using atomistic tools (see in particular Rudd and Belak, 2002).
However, the probabilistic approach is an interesting alternative,
as shown by Grady and Kipp (1979, 1980) and Denoual and Hild
(2000) for dynamic fragmentation of brittle materials, and more
recently by Molinari and Wright (2005) and Czarnota \coll\
(2006, 2008) for ductile spalling. In both cases, the purely
deterministic description of void growth is combined with a
stress-dependent probability of void nucleation, in the form of a
Weibull-like model. Czarnota \coll\ (2006, 2008) defined a
probability of nucleating new voids; in addition, Denoual and Hild
(2000) used a spatial distribution of crack nuclei among which new
cracks are activated. Void interactions are also treated in a
different fashion. Czarnota \coll\ (2006, 2008) used the
overall porosity to describe the weakening effect of the already
present voids, whereas Denoual and Hild (2000) considered
microcrack growth as a spatially bounded relaxation process that
inhibits nucleation in relaxed zones. In this respect, the degree
of coupling is stronger in the last approach.

\begin{figure}%[ht]
    \begin{center}
        \includegraphics[width=13cm]{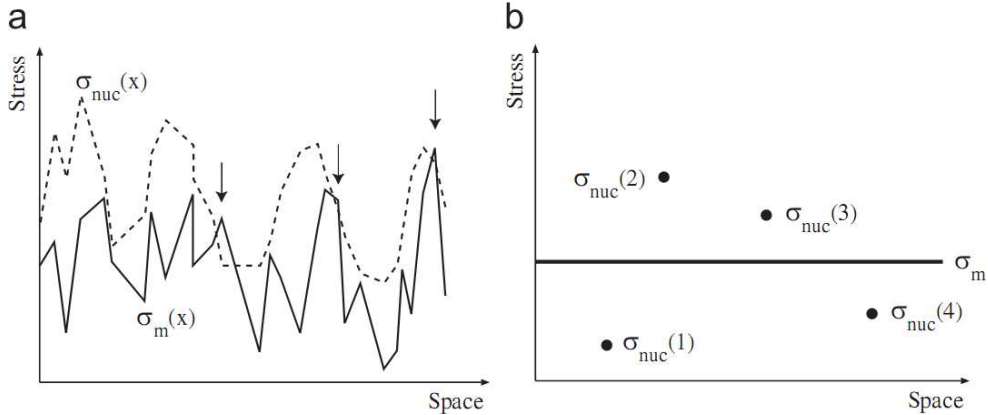}
    \caption{Simplifying assumptions of the nucleation model (nucleation conditions met at arrows).
        (a) Physical situation: applied tensile stress $\sigma_{\rm m}$ and nucleation level
        $\sigma_{\rm nuc}$ as continuous random fields. (b)
        Simplification: uniform applied stress, and field of nucleation thresholds split up
        into discrete sites of random locations and threshold values.}
    \label{fig:Figure2}
    \end{center}
\end{figure}

It is intended here to assess the relevance of
inhibition concepts for fragmentation (Mott, 1947; Grady and Kipp,
1980; Denoual and Hild, 2000) to analyze ductile spalling
processes. Rather than precisely describing joint nucleation,
growth processes and their couplings, this paper aims at setting
the fundamentals of the theory to demonstrate its potentialities
within the simplest possible theoretical framework.
Section \ref{sec:NucleationAndGrowth} shows how the deterministic
and probabilistic parts of the model are interlinked, and puts the
emphasis on inertial growth, which drives the inhibition process.
Section \ref{sec:RampLoading} presents an application to ramp
loading, generally agreed to be representative of the real loading
in the spall plane in the lack of any phase transition process,
and ends up with a closed-form solution of the whole problem. A
very simple overall damage model is proposed in Section
\ref{sec:ConstitutiveModel}, and yields an analytical expression
for the spall stress, i.e., the maximum tensile stress the material
can sustain during the whole spalling process. Through a thorough
examination of the experimental data of Nicollet \coll\
(2001), Roy (2003), and Bontaz-Carion and Pellegrini (2006) on
pure tantalum, the model is identified and discussed in Section
\ref{sec:TantalumExperiments}, and applied tentatively to other
materials in Section \ref{sec:AluminumAndMagnesium}.

%%%%%%%%%%%%%%%%%%%%%%%%%%%%%%%%%%%%%%%%%%%%%%%%%%%%%%%%%%%%%%%%%%%%%%%%%%%%%%%%%%%%%
%%%%%%%%%%%%%%%%%%%%%%%%%%%%%%%%%%%%%%%%%%%%%%%%%%%%%%%%%%%%%%%%%%%%%%%%%%%%%%%%%%%%%
%
\section{Nucleation and growth in ductile materials}
\label{sec:NucleationAndGrowth}
%
%%%%%%%%%%%%%%%%%%%%%%%%%%%%%%%%%%%%%%%%%%%%%%%%%%%%%%%%%%%%%%%%%%%%%%%%%%%%%%%%%%%%%
%
\subsection{Model outline}
\label{subsec:ModelOutline}
As introduced above, the physical process of nucleation and growth
during early stages of ductile spallation is complex. Wave
propagation induces transient macroscopic stress fields. At a
finer (mesoscopic) spatial scale, local fields experience
fluctuations due to the polycrystalline nature of the materials
considered (Fig.~\ref{fig:Figure2}a). When the local hydrostatic
stress $\mathbf{\sigma}_{\rm m}(\mathbf{x},t)$ exceeds some local
nucleation threshold $\sigma_{\rm nuc}(\mathbf{x})$, cavities are
nucleated and start to grow.

As shown by Roy and Villechaise (Roy, 2003), in pure tantalum
nucleation sites are primarily located at grain boundaries,
especially triple points.\footnote{This picture is valid when the
material is pure. When second-phase particles or precipitates are
present, the so-called heterogeneous nucleation processes take
place. The present paper focuses on the first mechanism.} Growing
cavities in turn induce relaxation zones in which local stresses
decrease, thus decreasing the probability of nucleating voids in
these zones, and out of which local stresses remain unaltered.
Hence, any volume element in which macroscopic stresses are
uniform prior to nucleation evolves into a volume containing
growing perturbed zones in an otherwise unperturbed uniformly
loaded matrix.
%%
%% PAGE 3
%%
%

According to Roy (2003), isolated voids remain spherical from very
small to very large sizes, implying that local fluctuations of
material properties do not seem to influence void growth. Hence,
a first simplification will consist in neglecting the effects of the
polycrystalline nature of the material of the matrix, and
therefore on macroscopic stresses.
We thus assume \emph{uniform loading}, in a pristine matrix
material that contains a \emph{random spatial distribution of
void nuclei} at which the elastic--plastic properties of matrix
material \emph{strongly fluctuate} around their bulk value
(Fig.~\ref{fig:Figure2}b).
Furthermore, the matrix is assumed perfectly plastic
hereafter. Neglecting temperature, viscosity and strain
hardening is performed for the sake of simplicity, and
can be relaxed in more detailed (future) analyses.
During further evolutions, activated voids are the only local
heterogeneities that will affect macroscopic stresses. In this
context, a voided volume is viewed as a matrix loaded by a uniform
hydrostatic tensile stress $\sigma_{\rm m}$, containing several
(possibly overlapping) perturbed zones.

Second, the joint effects of local stress fluctuations and of
local weaknesses are accounted for through a
\emph{stress-dependent nucleation probability}. It will be further
assumed that the inhibition phenomenon is total in strongly
relaxed zones. Hence, matrix stresses will be considered as the
only driving force for nucleation and growth.

Third, given the high level of triaxiality, as well as the
spherical shape of the voids observed by Roy (2003),
macroscopic shear stresses will be neglected, such that
$\sigma_{ij} = \sigma_{\rm m} \delta_{ij}$. From now on,
$\sigma_{\rm m}$ will simply be referred to as ``the stress.''

Growth drives the extension of relaxation zones, and thus the
inhibition process. The growth model must therefore be carefully
chosen. On the one hand, as stressed by Ortiz and Molinari (1992),
Wang and Jiang (1997), Roy (2003), Dragon and Trumel (2003),
Molinari and Wright (2005), and Czarnota \coll\ (2006), inertial
effects are overwhelmingly important. On the other hand,
elasticity should not be neglected, since it has a strong effect
on early growth (Denoual and Diani, 2002; Roy, 2003). Advantage
will be taken here of a simplified approach proposed by Roy
(2003), from the work of Forrestal and Luk (1998). This approach
shows that growth cannot take place if the macroscopic stress is
less than a \emph{cavitation} threshold, as shown by many authors
in the quasi-static case (Mandel, 1966; Hou and Abeyaratne, 1992;
Denoual and Diani, 2002).

We now proceed to assemble the above-listed ingredients. In a
pristine examination volume $V$ subjected to uniform stress
$\sigma_{\rm m}(t)$, we assume the number $N$ of \emph{active}
nucleation sites, of associated random nucleation stress
$\sigma_{\rm nuc}(\mathbf{x})$ where $\mathbf{x}$ is the site
location, to follow a point-Poisson distribution of intensity
$n_{\rm tot}$ (the average volume density of active sites).
The probability of finding $N$ active sites in $V$ is
\begin{eqnarray}
    \label{eqn:PoissonDistribution}
    P(N,V)=\frac{\left(n_{\rm tot}V\right)^{N}}{N!}\exp \left(-n_{\rm tot}V\right).
\end{eqnarray}
In the above definition, a nucleation site at location
$\mathbf{x}$ is said \emph{active} at $t$ (i.e., can potentially
nucleate a void) if $\sigma_{\rm m}(\tau)\geq \sigma_{\rm
nuc}(\mathbf{x})$ for any past time $0\leq \tau\leq t$. It will
effectively give birth to a void only if not inhibited (effects of
inhibition are dealt with in Section~\ref{subsec:InhibitionModel}).
Introduce then $\sigma_\mathrm{max}(t) = \max_{0 \leq \tau \leq t}
\sigma_{\rm m}(\tau)$, the maximum hydrostatic stress reached up
to time $t$. According to experimental findings (Roy, 2003), the
density of nucleated cavities is stress-dependent. This prompts us
to further write $P(N,V)$ in the form of the so-called
Weibull--Poisson law by taking (Gulino and Phoenix, 1991; Jeulin,
1991; Denoual and Hild, 2002)
\begin{eqnarray}
    \label{eqn:StressDependentDensity}
    n_{\rm tot}(t) = n_0 \left[ \frac{\langle \sigma_\textrm{max}(t) \rangle}{\sigma_0} \right]^m,
\end{eqnarray}
where $m$ is the Weibull modulus which characterizes the
\emph{scatter} in nucleation levels (weak scatter corresponds to a
high $m$ value, and conversely), $\sigma_0$ is a scale parameter
relative to a reference density $n_0$, and $\langle \star \rangle
$ are Macauley brackets that denote the positive part of $\star$.
In Eq.~(\ref{eqn:PoissonDistribution}), the product $n_{\rm
tot}V$ thus represents the average number of sites in $V$ where
$\sigma_{\rm m}$ has overcome the nucleation threshold.
Eq.~(\ref{eqn:StressDependentDensity}) indicates that the higher
$\sigma_\textrm{max}(t)$, the more nucleation sites are active. It
should be noted that a classical Weibull expression is retrieved
within the weakest link framework, see Appendix \ref{appendix:Weibull}.

%%
%% PAGE 4
%%
Since Eqs.~(\ref{eqn:PoissonDistribution}) and
(\ref{eqn:StressDependentDensity}) describe the probability of
activating $N$ sites in a pristine uniformly loaded volume $V$,
they also hold (with $V$ replaced by $V'$) in the \emph{uniformly
loaded part} $V'$ of a larger \emph{voided} volume, by definition
of $\sigma_{\rm m}$, and by the above assumption of total
inhibition. The volume $V'$ is found by subtracting from $V$ the
volume of inhibited zones, thus accounting for possible overlaps
between individual inhibition zones that grow out of each
activated site. Since inhibition is related to stress relaxation,
$V'$ depends on the growth model, which is addressed now.
%
%%%%%%%%%%%%%%%%%%%%%%%%%%%%%%%%%%%%%%%%%%%%%%%%%%%%%%%%%%%%%%%%%%%%%%%%%%%%%%%%%%%%%
%
\subsection{A simplified growth model}
\label{sec:simplified}
\begin{figure}[ht]
    \centering
    \includegraphics[width=6.5cm]{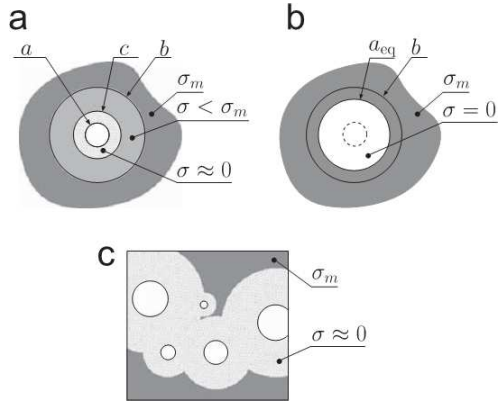}
 \caption{Equivalent hollow sphere model. (a) Real elasto-plastic hollow sphere. (b) Simplified representation of equivalent elastic energy. (c) Schematic representation of overlapping relaxed zones. See Section~\ref{sec:ConstitutiveModel}
    for a discussion of (b) and (c).}
    \label{fig:Figure3}
\end{figure}
Cavity nucleation can be understood as a bifurcation process in
the sense of Hou and Abeyaratne (1992). Once nucleated, any new
cavity starts to grow. As shown, for example, by Hopkins (1960),
Hunter and Crozier (1968), Glennie (1972), or Roy (2003), an
isolated growing cavity of radius $a(t)$ can be seen as an
expanding volume bounded by an elastic relaxation wave at radius
$b(t)$. This volume consists in an outer elastic zone, and an
inner elastic--plastic region, separated by an evolving
boundary of (``plastic'') radius $c(t)$
(Fig.~\ref{fig:Figure3}a). Both regions are referred to as
``the matrix'' hereafter.
Denoual and Diani (2002) and Tonks \coll\ (2001) showed that the
early growth can be decomposed into three distinct phases. The
first one is essentially elastic, until the hydrostatic stress
reaches a ``cavitation threshold" (see below). There, bulk elastic
energy release induces a violent elastic--plastic expansion of the
cavity, until the third phase of stationary expansion is
established.

\label{subsec:SimplifiedGrowthModel}
\begin{figure}[ht]
    \centering
    \includegraphics[width=12cm]{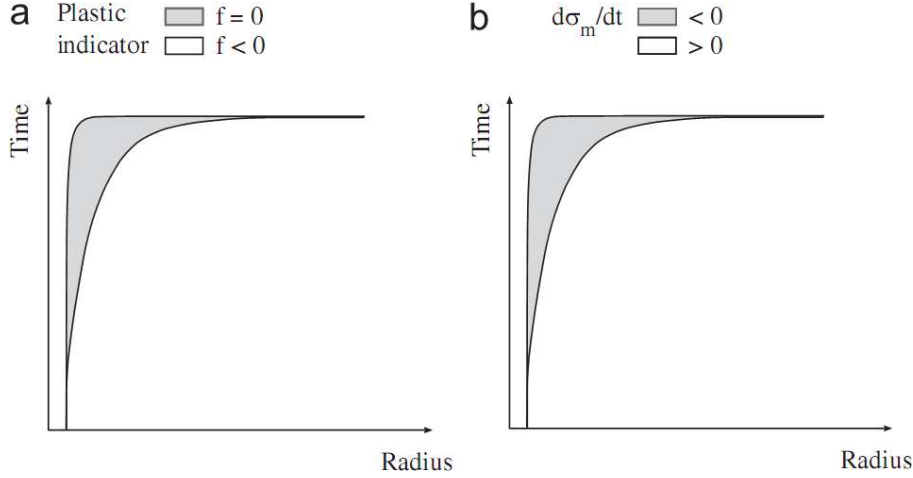}
    \caption{Space-time diagram for a hollow sphere (initial outer radius: $b=12.5$ $\mathrm{\mu}$m, initial inner radius
$a = 0.58$ $\mathrm{\mu}$m) submitted to a stress rate of $2$
GPa/$\mathrm{\mu}$s on the outer radius. The material
parameters are those of tantalum (Table \ref{table:Table1}). The plastic zone coincides with the region in which the
local pressure decreases, in spite of an overall pressure increase. (a) Plastic zone evolution and (b) stress rate indicator.}
    \label{fig:Figure4}
\end{figure}

Strong relaxation occurs inside the elastic--plastic zone. This is
illustrated numerically, by submitting a hollow tantalum sphere of
initial outer and inner radii of 12.5 and 0.58~$\mu$m, respectively (i.e., an initial porosity of $10^{-4}$) to a
hydrostatic stress ramp of 2 GPa $\mathrm{\mu}$s$^{-1}$ applied on
the outer boundary. Fig.~\ref{fig:Figure4} shows the space--time
domain where the matrix is yielding (Fig.~4a), and
that of varying $\sigma_{\rm m}$ (Fig.~4b). It is
seen that $\sigma_{\rm m}$ decreases inside the plastic zone
although the applied stress keeps increasing. Plastic zones can
thus be seen as (and identified to) \emph{inhibition zones} for
further void nucleation, and this is exploited in the next
section. This shows that unlike previous works (Wu \coll, 2003;
Molinari and Wright, 2005; Czarnota \coll, 2006), it is not
sufficient to establish a link between the macroscopic stress and
the cavity radius $a$, namely the link between these quantities
and $c$ must be known as well.

Roy \coll\ (2002) and Roy (2003) checked numerically that the
cavitation stress is independent of the macroscopic strain rate,
and that the transient regime is brief. Accordingly, and since
this allows for closed-form solutions, a purely stationary model
is used here, with $a(t)=\dot{a} t$, $c(t)=\dot{c} t$, where
$\dot{a}$ and $\dot{c}$ are constant growth velocities. The
approach used (detailed in Appendix~\ref{appendix:ForrestalLuk})
is adapted from the work of Forrestal and Luk (1988), itself
derived from earlier works dealing with isolated cavity growth
under internal hydrostatic stress (Hopkins, 1960; Hunter and
Crozier, 1968).

Thus, for an isolated cavity in an infinite medium subjected to a
remote tensile stress $\sigma_{\rm m}(t)$, an implicit equation
links $\dot{c}$ and $\sigma_{\rm m}(t)$ to $\dot{a}$
(Eqs.~(\ref{eqn:CdotVersusAdotExact}) and
(\ref{eqn:PressureVersusAdotExact})). A first-order expansion
valid in the low stress rate regime (assuming $\dot{c} \ll c_P$
\emph{and} $\dot{a}$ $\ll c_P$, where $c_P=\sqrt{K/\rho_0}$ is the
so-called plastic velocity (Zel'dovich and Raizer, 2002), $K$ is
the bulk modulus, and $\rho_0$ is the reference density), then
provides the additional proportionality relationship
\begin{eqnarray}
    \label{eqn:CdotVersusAdot}
    \dot{c} = \beta^{-1/3}\, \dot{a},
\end{eqnarray}
where $\beta$ is defined by Eq.~(\ref{eqn:CdotVersusAdotLinear})
in terms of $K$, $\mu$ the shear modulus, and $Y$ the yield
stress. For most materials $\beta\ll 1$. In turn, a similar
first-order expansion provides relationship
(\ref{eq:cavitationbehavior}), namely,
\begin{eqnarray}
    \label{eqn:AdotVersusPressure}
    \dot{a}=\dot{a}_0\langle\sigma_{\rm m}/\sigma_{\rm
cav}-1\rangle^{1/2}
\end{eqnarray}
between the void growth velocity and the applied tensile stress,
where $\sigma_{\rm cav}$ is the cavitation threshold, and where
$\dot{a}_0$ is a characteristic void growth velocity in the
material. Both quantities depend only on $K$, $\mu$ and $Y$, with
$\dot{a}_0$ depending on $\rho_0$ as
%%
%% PAGE 5
%%
well (see Appendix~\ref{appendix:ForrestalLuk} for explicit expressions 
of these quantities). Combining Eqs.~(\ref{eqn:CdotVersusAdot}) and
(\ref{eqn:AdotVersusPressure}) yields
\begin{eqnarray}
    \label{eqn:CdotVersusPressure} \dot{c}=k\, c_P\,\langle\sigma_{\rm
m}/\sigma_{\rm cav}-1\rangle^{1/2},
\end{eqnarray}
where $k\equiv\beta^{-1/3}\dot{a}_0/c_P$ is a numerical
coefficient ($k\, c_P$ being a characteristic growth velocity of
the plastic region). For Al, Cu and Ta, $k$ varies between 0.3 and
0.5.

Eq.~(\ref{eqn:CdotVersusPressure}) constitutes a particular
instance of the more general class of threshold-like expressions
\begin{eqnarray}
    \label{eqn:CdotVersusPressureGeneral}
    \dot{c}=k\,c_P \langle \sigma_{\rm m}/\sigma_{\rm cav}-1\rangle^{\alpha},
\end{eqnarray}
where $\alpha \geq 0$ is a stress-sensitivity exponent, and where
the nucleation stress $\sigma_{\rm nuc}$ of Fig.~\ref{fig:Figure2}
is identified to the cavitation threshold $\sigma_{\rm cav}$. No
significant growth of the microvoid population should occur unless
cavitation conditions are met. This general expression covers the
present case, as well as the ``quasi-brittle'' case for which
$\alpha=0$ (Denoual \coll, 1997). In the case of monotonically
increasing loading $\sigma_{\rm m}(t)$, upon integrating
(\ref{eqn:CdotVersusPressureGeneral}) over time we obtain $c(t)$
in the form
\begin{eqnarray}
    \label{eqn:CMonotonous}
    c(t)=C(t-t_{\rm nuc}),\qquad (t>t_{\rm nuc}),
\end{eqnarray}
where $C$ is some function and $t_{\rm nuc}$ the
nucleation time obtained as a solution to
\begin{eqnarray}
\label{eq:mapping} \sigma_{\rm m}(t_{\rm nuc})=\sigma_{\rm cav}.
\end{eqnarray}
%
%%%%%%%%%%%%%%%%%%%%%%%%%%%%%%%%%%%%%%%%%%%%%%%%%%%%%%%%%%%%%%%%%%%%%%%%%%%%%%%%%%%%%
%
\subsection{Elementary cell assembly}
\subsubsection{Dynamic inhibition model}
\label{subsec:InhibitionModel} So far, we described the behavior
of \emph{isolated} cavities only, in a deterministic way. The
\emph{collective} behavior of the population of voids is now
considered. Henceforth, overlined quantities are used for
macroscopic variables that represent statistical (or more
phenomenological) averages of their microscopic counterparts.

The intrinsic probabilistic nature of the nucleation and growth
process should be embodied in some random variability of the local
elastic and plastic properties of the material $Y$, $\mu$, $K$ and
$\rho_0$. Eq.~(\ref{eq:mapping}) shows that under some
prescribed time-dependent loading, a random set $\{\sigma_{\rm
cav}\}$ of cavitation or generic nucleation thresholds (see
Fig.~\ref{fig:Figure2}) can be mapped to a random set $\{t_{\rm
nuc}\}$ of nucleation times. Randomness in the process is thus
introduced through the following crucial bold assumption that
emphasizes the part played by nucleation times, namely, in
Eq.~(\ref{eqn:CMonotonous}) the nucleation time $t_{\rm nuc}$,
which physically depends on the above material parameters \emph{and} on
the local loading, will be considered as a random variable,
whereas material parameters, and parameters that define the field
loading function, will be considered as ``averaged'' ones whenever
they enter the definition of the function $C$ itself.

Section \ref{sec:simplified} substantiates the identification
between plastic regions and zones of total nucleation inhibition.
Accordingly, the inhibition volume $V_{\rm inh}$ associated to an
\emph{isolated} cavity is taken hereafter proportional to the
plastic radius $c$ to the third power
\begin{eqnarray}
    \label{eqn:ObscurationVolume}
    V_{\rm inh}=V_{\rm inh}(t-t_{\rm nuc})=S\,c^{3},
\end{eqnarray}
where $S$ is a shape parameter, and the functional time dependence
of $V_{\rm inh}$ stems from Eq.~(\ref{eqn:CMonotonous}).
\begin{figure}[ht]
    \centering
    \includegraphics[width=13cm]{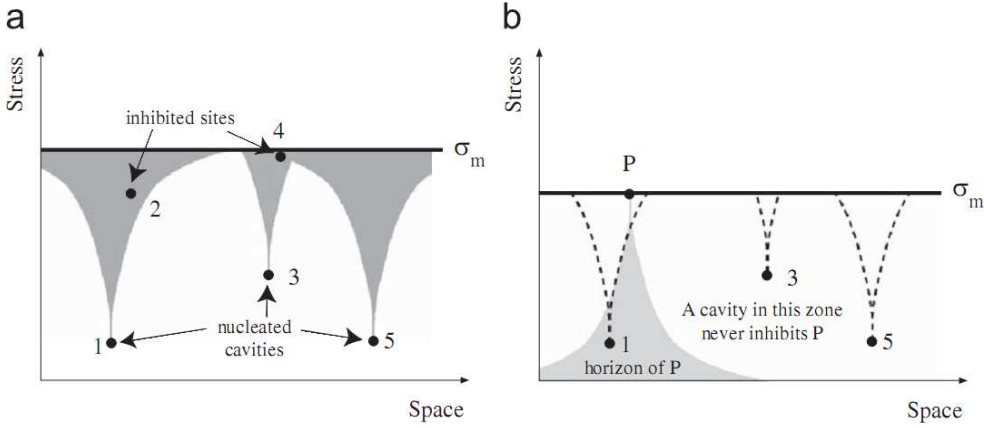}
    \caption{
    Inhibition and horizon concepts. (a) Inhibition phenomena. In grey are relaxed zones where void nucleation is inhibited. (b) Horizon of a site $P$. Any active site in the grey zone
    inhibits further cavity nucleation at $P$.}
    \label{fig:Figure5}
\end{figure}

New voids will nucleate from active nucleation sites (in the sense
of Sec.~\ref{subsec:ModelOutline}) only if they do not belong to
\emph{any} relaxed zone produced by previously nucleated growing
voids, as depicted in Fig.~5a. Thus $n_{\rm nuc}$,
the volume density of the centers of nucleated voids, is related to
$n_{\rm tot}$ defined in Eq.~(\ref{eqn:StressDependentDensity}) by
\begin{eqnarray}
    \label{eqn:DifferentialEquation}
    \frac{\mathrm{d} n_{\rm nuc}}{\mathrm{d} t} = \left(1-P_{\rm inh}\right) \frac{\mathrm{d}
    n_{\rm tot}}{\mathrm{d}t}
\end{eqnarray}
with $n_{\rm nuc}(0) = n_{\rm tot}(0) = 0$. This equation, which
implements inhibition effects in the model, involves the
inhibition
%%
%% PAGE 6
%%
probability (identified with an overall volume fraction
of inhibited regions)
\begin{eqnarray}
    \label{eqn:ProbabilityOfObscuration}
    P_{\rm inh}(t)=1-\exp \left[ -\overline{V}_{\rm inh}(t)
    n_{\rm tot}\left\{\sigma_{\rm m}(t)\right\} \right],
\end{eqnarray}
where $\overline{V}_{\rm inh}$, the \emph{mean} volume of the
inhibition zone, is defined by
\begin{eqnarray}
    \label{eqn:AverageObscurationVolume}
    \overline{V}_{\rm inh}(t) n_{\rm tot} \left\{ \sigma_{\rm m}(t) \right\} = \int_{0}^{t}
    V_{\rm inh}(t - \tau) \frac{\mathrm{d} n_{\rm tot}}{\mathrm{d} \tau} \left\{
         \sigma_{\rm m}(\tau) \right\} \mathrm{d} \tau.
\end{eqnarray}
Eqs.~(\ref{eqn:ProbabilityOfObscuration}) and
(\ref{eqn:AverageObscurationVolume}) (Denoual \coll, 1997), which
originate from the Poisson hypothesis, Eq.~(\ref{eqn:PoissonDistribution}), are derived in Appendix~\ref{appendix:ObscurationProbability}, which makes clear that Eq.~(\ref{eqn:ProbabilityOfObscuration}) \emph{accounts for the
overlaps} of inhibition zones (the derivation uses the
\emph{horizon} concept described in Fig.~5b, which
constitutes another way to look at the inhibition process). From
the point of view of mathematical morphology, this model
constitutes an instance of a Boolean islands model (Jeulin and
Jeulin, 1981; Serra, 1982). Also, in the context of isothermal
diffusive phase transformations, the three latter equations are
central to the Kolmogorov--Johnson--Mehl--Avrami (KJMA) kinetic
theory of nucleation and growth (Kolmogorov, 1937; Johnson and
Mehl, 1939; Avrami, 1941). Eqs.~(\ref{eqn:ProbabilityOfObscuration}) and 
(\ref{eqn:AverageObscurationVolume}) are valid for any density
$n_{\rm tot}$ and any shape of interaction zones of volume $V_{\rm
inh}$. The present framework is thus adaptive to incorporate many
different inhibition phenomenologies. In particular, the same
approach can be used to analyze dynamic fragmentation of brittle
materials (Grady and Kipp, 1979, 1980; Denoual and Hild, 2000, 2002). In that case, inhibition is induced by stress relaxation around propagating cracks, as was also the case for the shell fragmentation problem studied by Mott (1947).

Bearing in mind the particular time dependence of $V_{\rm inh}$ in
Eq.~(\ref{eqn:ObscurationVolume}), it is observed that the
time-integration in Eq.~(\ref{eqn:AverageObscurationVolume}) is
over the nucleation time. According to our above hypothesis of
considering the nucleation time as a random variable,
Eq.~(\ref{eqn:AverageObscurationVolume}) indicates that its
associated probability density at time $t$ imposed by the
Weibull--Poisson process~(\ref{eqn:StressDependentDensity}) is
(with $\tau\geq 0$)
\begin{eqnarray}
\mathcal{P}(t_{\rm nuc}=\tau;t)=\frac{\theta(t-\tau)}{n_{\rm
tot}\left\{\sigma_{\rm m}(t) \right\}}\frac{\mathrm{d} n_{\rm
tot}}{\mathrm{d} \tau}\left\{\sigma_{\rm m}(\tau) \right\},
\end{eqnarray}
where $\theta$ is the Heaviside step function.

Finally, an expression for the average void volume fraction $f$ in
the examination volume is obtained as follows. Eq.~(\ref{eqn:CdotVersusAdot}) implies, via $a=\beta^{1/3}\, c$, the
following proportionality relationship between the individual
cavity volume $V_{\rm cav}\propto a^3$ and the corresponding
inhibition volume $V_{\rm inh}\propto c^3$
\begin{eqnarray}
    \label{eqn:CavityVolumeVersusObscurationVolume}
    V_{\rm cav} = \beta\,V_{\rm inh}.
\end{eqnarray}
Using Eq.~(\ref{eqn:AverageObscurationVolume}), the average
cavity volume $\overline{V}_{\rm cav}$ follows as
\begin{eqnarray}
    \overline{V}_{\rm cav}(t) = \beta\, \overline{V}_{\rm inh}(t).
\end{eqnarray}
Since the individual voids and inhibition zones are of same
centers, they obey the same statistics. The porosity $f$ is thus
\begin{eqnarray}
    \label{eqn:Porosity}
    f(t)=1-\exp \left[ -\overline{V}_{\rm cav}(t)
    n_{\rm tot}\left\{\sigma_{\rm m}(t)\right\} \right]
\end{eqnarray}
and simply relates to the inhibition probability by
\begin{eqnarray}
    \label{eq:fpinh}
    f=1-(1-P_{\rm inh})^{\beta}.
\end{eqnarray}
This relationship is illustrated by Fig.~3c, interpreting in the present context white zones as voids of overall volume fraction $f$, and dotted zones as inhibited zones of overall volume fraction $P_{\rm inh}$.
%%
%% PAGE 7
%%
\subsubsection{Application to ramp loading}
\label{sec:RampLoading}

In general the number of nucleated
cavities must be computed numerically. The
nucleation Eq.\ (\ref{eqn:DifferentialEquation}) involves the
matrix stress in the non-inhibited zones, $\sigma_{\rm m}$.
The link with the overall stress is given in Section~\ref{sec:ConstitutiveModel}. The computation is particularly
simple for the particular case of ramp-stress loading
$\sigma_{\rm m}=\dot{\sigma}\, t$ with constant stress-rate
$\dot{\sigma}$ that yields a {\em closed-form} solution of practical interest for experimental analyses. Upon integrating
Eq.~(\ref{eqn:CdotVersusPressureGeneral}) over time, and
introducing the nucleation time $t_{\rm nuc}\equiv \sigma_{\rm
cav}/\dot{\sigma}$ according to the first paragraph of
Section~\ref{subsec:InhibitionModel}, the individual inhibition
volume~(\ref{eqn:ObscurationVolume}) reads
\begin{eqnarray}
    \label{eqn:ObscurationVolumeRamp}
    V_{\rm inh}=S \left[ \frac{k\,
    c_P}{\alpha + 1} \left( \frac{\dot{\sigma}}{
    \sigma_{\rm cav}}\right)^{\alpha} \left( t - t_{\rm nuc} \right)^{\alpha + 1}
    \right]^3
\end{eqnarray}
for $t>t_{\rm nuc}$, and zero otherwise. The corresponding cavity
volume $V_{\rm cav}$ follows from
Eq.~(\ref{eqn:CavityVolumeVersusObscurationVolume}).

At this stage, it proves useful to introduce a dimensionless flaw
density $\widetilde{n}=n/n_c$, time $\widetilde{t}=t/t_c$, volume
$\widetilde{V}= V/V_c$ and stress $\widetilde{\sigma}_{\rm
m}=\sigma_{\rm m}/\sigma_{\mathrm{c}}$. Two ways of defining those
dimensionless quantities are relevant here. Both are based on the
condition
\begin{eqnarray}
    \label{eqn:NonDimensional}
    n_c V_c = 1\qquad (t=t_c)
\end{eqnarray}
that expresses the fact that some characteristic volume $V_c$
contains {\em on average} one site at time $t_c$.

Computing $P_{\rm inh}$ requires identifying $V_c$ with the
\emph{inhibition} volume, whereby the above condition reads
\begin{eqnarray}
    \label{eqn:NonDimensionalObscurationVolume}
    n_{\rm ci} V_{\rm ci} = 1,\quad
    n_{\rm ci} = n_{\rm tot}[\sigma_{\rm m} (t_{\rm ci})],\quad
    V_{\rm ci} = V_{\rm inh}(t_{\rm ci}),
\end{eqnarray}
where the subscript $ci$ denotes characteristic quantities
associated to inhibition. A characteristic stress is defined by
$\sigma_{\rm ci} = \dot{\sigma}\, t_{\rm ci}$. From
Eqs.~(\ref{eqn:StressDependentDensity}) and
(\ref{eqn:NonDimensionalObscurationVolume}), the characteristic
parameters follow as
\begin{eqnarray}
    \label{eqn:CharacteristicParametersObscurationVolume}
    t_{\rm ci} &=& \left[\frac{(\alpha+1)^3 \sigma_{0}^{m}
    \sigma_{\rm cav}^{3 \alpha}}{n_0\,(k
    c_P)^{3}\,S\,\dot{\sigma}^{m+3\alpha}} \right]^{1/[m+3(\alpha+1)]}, \nonumber \\
    V_{\rm ci} &=& \left[ \frac{
    k\,c_P\,S^{1/3}\,\sigma_{0}^{\alpha + 1}}{(\alpha + 1) n_{0}^{(\alpha + 1)/m}
    \sigma_{\rm cav}^\alpha\,\dot{\sigma}} \right]^{3m/[m + 3 (\alpha + 1)]},\nonumber \\
    \sigma_{\rm ci} &=& \left[ \frac{(\alpha + 1)^3\sigma_{0}^{m}
    \sigma_{\rm cav}^{3 \alpha} \dot{\sigma}^{3}}{n_0\,(k
    c_P)^{3}\,S} \right]^{1/[m+3(\alpha+1)]}.
\end{eqnarray}
Upon carrying out the integration in
Eq.~(\ref{eqn:AverageObscurationVolume}),
Eq.~(\ref{eqn:ProbabilityOfObscuration}) reads
\begin{eqnarray}
    \label{eqn:ProbabilityOfObscurationClosedForm}
    P_{\rm inh}=1-\exp \left[-B\bigl(m,3(\alpha+1)\bigr) \widetilde{t}^{\,m+3(\alpha+1)}
    \right],
\end{eqnarray}
where $B$ is a modified Euler function of the first kind
\begin{eqnarray}
    \label{eqn:ModifiedEuler}
    B(p,q) = p \int_0^{1} t^{p-1} (1-t)^{q} \mathrm{d}t =
    \frac{\Gamma(p+1) \Gamma(q+1)}{\Gamma(p+q+1)},
\end{eqnarray}
and the closed-form solution of
Eq.~(\ref{eqn:DifferentialEquation}) yields
\begin{eqnarray}
    \label{eqn:AnalyticalSolutionRamp}
    \widetilde{n}_{\rm nuc}\left(\widetilde{t}\right)=\frac{m \;B\bigl(m,3(\alpha+1)\bigr)^{-m/[m+3(\alpha+1)]}}{m+3(\alpha+1)}\,
    \gamma\left( \frac{m}{m+3(\alpha+1)}\; ,
    B\bigl(m,3(\alpha+1)\bigr) \widetilde{t}^{\,m+3(\alpha+1)} \right),
\end{eqnarray}
where $\gamma$ is the incomplete gamma function
\begin{eqnarray}
    \label{eqn:IncompleteGamma}
    \gamma(p,x) = \int_0^{x} t^{p-1} e^{-t} \mathrm{d}t
\end{eqnarray}
so that $\gamma(p,x \rightarrow + \infty) = \Gamma(p)$.
\begin{figure}[ht]
    \centering
    \includegraphics[width=8.5cm]{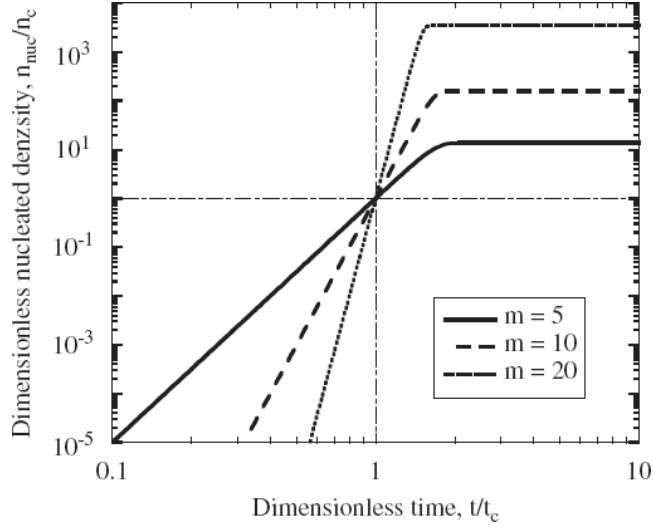}
     \caption{Dimensionless nucleated density $\widetilde{n}_{\rm nuc}$,
     Eq.~(\ref{eqn:AnalyticalSolutionRamp}), vs.~dimensionless time
     $\widetilde{t}$ for three different Weibull moduli $m$ when $\alpha=1/2$
     ($n_{\rm c}$ denotes either $n_{\rm ci}$ or $n_{\rm cc}$).}
    \label{fig:Figure6}
\end{figure}
Eq.~(\ref{eqn:AnalyticalSolutionRamp}) is the \emph{exact}
solution to Mott's problem (1947) extended to three-dimensional cases with an initial flaw density modeled by a power law function.
Fig.~\ref{fig:Figure6} shows the change of the dimensionless
density $\widetilde{n}_{\rm nuc}$ with the dimensionless time
$\widetilde{t}$. At early times $\widetilde{t} < 1$, virtually no
inhibition is observed, i.e., $P_{\rm inh} \approx 0$ and
$\widetilde{n}_{\rm nuc} \approx \widetilde{n}_{\rm tot}$.
Conversely, at late times $\widetilde{t} \gg 1$, $P_{\rm inh}
\approx 1$ and saturation occurs. The higher the Weibull modulus
$m$, the higher the density at saturation (Fig.~\ref{fig:Figure7}).

\begin{figure}[ht]
    \centering
    \includegraphics[width=8.5cm]{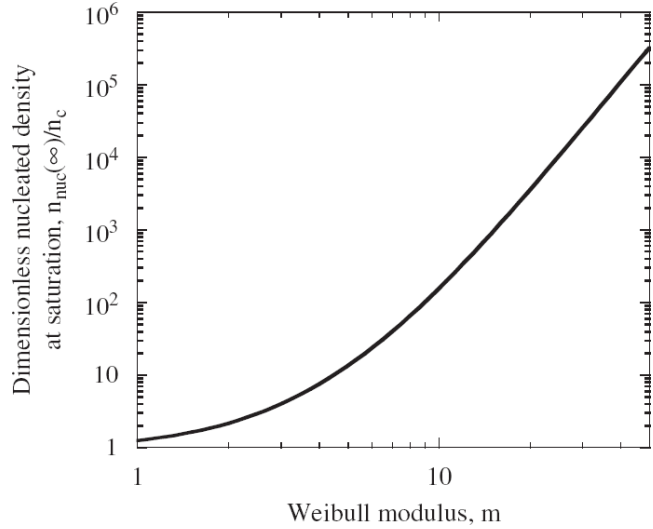}
     \caption{Dimensionless nucleated density at saturation,
     $\widetilde{n}_{\rm nuc}(\infty)$, vs.~modulus $m$ when $\alpha=1/2$
     ($n_{\rm c}$ denotes either $n_{\rm ci}$ or $n_{\rm cc}$).}
    \label{fig:Figure7}
\end{figure}

Computing $f$ instead requires identifying $V_c$ with the
\emph{void} volume, whereby the characteristic parameters obey
\begin{eqnarray}
    \label{eqn:NonDimensionalCavityVolume}
    n_{\rm cc} V_{\rm cc} = 1,\quad
    n_{\rm cc} = n_{\rm tot}[\sigma_{\rm m}(t_{\rm cc})],\quad
    V_{\rm cc} = V_{\rm cav}(t_{\rm cc}),
\end{eqnarray}
where the subscript ``cc'' denotes characteristic quantities
associated to cavities. Similarly, the characteristic stress is
defined by $\sigma_{\rm cc} = \dot{\sigma}\, t_{\rm cc}$. Then,
Eqs.~(\ref{eqn:CavityVolumeVersusObscurationVolume}) and
(\ref{eqn:NonDimensionalCavityVolume}) provide
\begin{eqnarray}
    \label{eqn:CharacteristicParametersCavityVolume}
    t_{\rm cc} = t_{\rm ci} \beta^{-1/[m+3(\alpha+1)]},\quad
    V_{\rm cc} = V_{\rm ci} \beta^{m/[m+3(\alpha+1)]},\quad
    \sigma_{\rm cc} = \sigma_{\rm ci} \beta^{-1/[m+3(\alpha+1)]},
\end{eqnarray}
and the overall porosity $f$, Eq.~(\ref{eqn:Porosity}), takes on the
form
\begin{eqnarray}
    \label{eqn:PorosityClosedForm}
    f=1-\exp \left[ -B\bigl(m,3(\alpha+1)\bigr)
    \widetilde{t}^{m+3(\alpha+1)} \right].
\end{eqnarray}
Remark that Eq.~(\ref{eqn:CharacteristicParametersCavityVolume})
follows from replacing $k$ by $k \beta^{1/3}$ in
Eq.~(\ref{eqn:CharacteristicParametersObscurationVolume}).
The only quantitative difference between
Eqs.~(\ref{eqn:PorosityClosedForm}) and
%%
%% PAGE 8
%%
(\ref{eqn:ProbabilityOfObscurationClosedForm}) resides in the
definition of the characteristic parameters (i.e., $\widetilde{t} =
t/t_{\rm ci}$ for inhibition and $\widetilde{t} = t/t_{\rm cc}$
for cavities). These results are exploited below in the framework
of a simplified constitutive model.
%
%%%%%%%%%%%%%%%%%%%%%%%%%%%%%%%%%%%%%%%%%%%%%%%%%%%%%%%%%%%%%%%%%%%%%%%%%%%%%%%%%%%%%
%%%%%%%%%%%%%%%%%%%%%%%%%%%%%%%%%%%%%%%%%%%%%%%%%%%%%%%%%%%%%%%%%%%%%%%%%%%%%%%%%%%%%
%
\section{A simplified constitutive model}
\subsection{Homogenization approach for dynamic loadings}
\label{sec:ConstitutiveModel}

In usual homogenization approaches to the computation of the
overall constitutive law of disordered porous media, some void
spatial distribution is prescribed in advance, \emph{all} voids
being by hypothesis in mutual long-range elastic interaction, and
the homogenization problem amounts to finding suitable
approximation schemes for these interactions. Such approaches
quite generally provide estimates of stress fluctuations in the
matrix (due to pore elastic interactions), which can be considered
as evenly spread in the latter. In stark contrast with this
situation, the dynamical impact conditions considered here consist
in loading a pristine matrix with a uniform stress state
$\sigma_{\rm m}$ in the first place, this initial state being
perturbed afterwards by relaxation waves originating from
nucleated growing voids. As a consequence, stress fluctuations in
the matrix are more localized (at least until significant overall
relaxation is achieved through some ``percolation'' of the relaxed
zones), and it should be clear that standard homogenization
techniques ought not be straightforwardly transposed to this case.

The following alternative two-step approach is adopted instead,
motivated by the elastic decoupling of the voids in the first
stages of the spall process. In a first step, the elementary
voided elastic--plastic cell of radius $b$, with void radius $a$,
in which the stress is heterogeneous but equal to $\sigma_{\rm m}$
on its boundary (Fig.~3a), is replaced by an
equivalent cell of radius $b$ containing a fictitious void of
radius $a_{\rm eq}$ (region of null stress), outside which the
stress is \emph{uniform} and equal to $\sigma_{\rm m}$
(Fig.~3b). The volume fraction of fictitious void
in the equivalent cell being written $\delta(c/b)^3$, where
$\delta$ is an unknown proportionality
%%
%% PAGE 9
%%
constant, it is proposed here to compute $\delta$ by requiring the elastic energy densities in the real and fictitious systems to be equal. The equation for $\delta$ thus reads
\begin{eqnarray}
    \label{eq:fordelta}
    \frac{1}{2}\langle\sigma:\mathbb{C}^{-1}:\sigma\rangle_{\rm cell}
    =\left[1-\delta(c/b)^3\right]\frac{1}{2}\frac{\sigma_{\rm
    m}^2}{K},
\end{eqnarray}
where $\langle\cdot\rangle_{\rm cell}$ denotes a volume average
over the elementary cell and where $\mathbb{C}$ is the usual
(isotropic) tensor of elastic moduli built on $K$ and $\mu$. The
l.h.s.~of Eq.~(\ref{eq:fordelta}), which involves microscopic
hydrostatic \emph{and} shear stress components, can be computed
using the stress of the exact solution for the real
elastic--plastic cell, derived in Appendix
\ref{appendix:ForrestalLuk}. Since this solution also provides $a$
and $b$ in terms of $\sigma_{\rm m}$, the outcome is an expression
of $\delta$ as a function of $\sigma_{\rm m}$. The associated
fictitious void volume $V_{\rm eq}=S a_{eq}^3=\delta V_{\rm inh}$
is then introduced (it is recalled that $V_{\rm inh}=S c^3$).

Since fictitious voids obey the same point-Poissonian statistics
as real ones, the same token that was used to relate $f$ to
$P_{\rm inh}$ given the relationship between $V_{\rm cav}$ and
$V_{\rm inh}$ in Sec.~\ref{subsec:InhibitionModel}, can be re-used
here to relate $P_{\rm inh}$ to an overall fictitious porosity
$f_{\rm eq}$ given the above relationship between $V_{\rm inh}$
and $V_{\rm cav}$. This second step provides the macroscopic
relationship analogous to Eq.~(\ref{eq:fpinh})
\begin{eqnarray}
    1-f_{eq}=(1-P_{\rm inh})^\delta.
\end{eqnarray}
In the macroscopic equivalent system, the stress outside the
fictitious voids is now \emph{homogeneous everywhere}, equal to
$\sigma_{\rm m}$ (Fig.~\ref{fig:Figure3}c). Hence the
ex\-pres\-sion of the ma\-cros\-co\-pic stress
$\overline{\sigma}_{\rm m}$ in terms of $\sigma_{\rm m}$ reads
\begin{eqnarray}
   \label{eq:constit}
    \overline{\sigma}_{\rm m}=(1-f_{eq})\sigma_{\rm m}=(1-P_{\rm inh})^\delta\sigma_{\rm
    m},
\end{eqnarray}
where $P_{\rm inh}$ and $\delta$ depend on $\sigma_{\rm m}$. In
this relation $f_{\rm eq}$ plays the part of an overall damage
variable $\overline{D}$ in the standard relation
$\overline{\sigma}_{\rm m}=(1-\overline{D})\sigma_{\rm m}$ of
damage theory (Lemaitre and Chaboche, 1990). To emphasize this
connection the notation $\overline{D}=f_{\rm eq}$ is used from now
on.

The computation of $\delta$ from the solution of Appendix
\ref{appendix:ForrestalLuk} is quite involved. Besides,
Eq.~(\ref{eq:fordelta}) is not free from arbitrariness since
other energetic equivalences could be proposed that explicitly
involve an additional kinetic energy term as proposed
by Wang and Jiang (1994) and Molinari and Mercier (2001). Bearing
in mind the present exploratory purpose, a pragmatic and
simplified approach is preferred that consists in considering only
the limiting cases $\delta=1$, whereby $\overline{D}=P_{\rm inh}$,
and $\delta=\beta$ whereby $\overline{D}=f$. These limits,
respectively, provide upper and lower ``pseudo-bounds'' (if not
rigorous ones) to $\overline{D}$. The former assumes that
\emph{relaxation is total in the plastic zone}, and neglects
elastic relaxation, such that the equivalent volume is the plastic
zone volume. The latter \emph{neglects any relaxation}, such that
the equivalent volume is the void volume. The relevance of these
``bounds'' is established below in Sec.~\ref{subsec:Validation} by
comparison to experimental results.

\begin{figure}[ht]
    \centering
    \includegraphics[width=8.5cm]{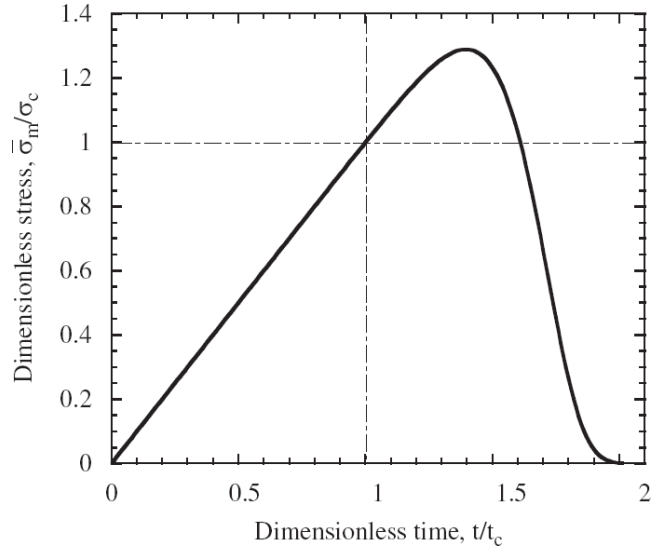}
    \caption{Dimensionless macroscopic stress provided by
    Eq.~(\ref{eq:constit}), vs.~dimensionless time when $m = 8$ and $\alpha=1/2$ ($\sigma_{\rm c}$ denotes either $\sigma_{\rm ci}$ or $\sigma_{\rm cc}$).}
    \label{fig:Figure8}
\end{figure}

For ramp loading, the constants required to write down in
dimensionless form the equations of type (\ref{eq:constit}) that
stem from each ``bound'' have been worked out in
Sec.~\ref{sec:RampLoading}. An example of the dimensionless
macroscopic stress $\overline{\sigma}_{\rm m}/\sigma_c$ as a
function of the dimensionless time $\widetilde{t}$, which reduces
to the same master curve for both ``bounds'', is displayed in
Fig.~\ref{fig:Figure8}.

Completing the above approach in order to arrive at a full
constitutive relationship between $\overline{\sigma}_m$ and the
macroscopic strain $\overline{\varepsilon}_m$ requires making
additional assumptions, and is not needed here.
%
%%%%%%%%%%%%%%%%%%%%%%%%%%%%%%%%%%%%%%%%%%%%%%%%%%%%%%%%%%%%%%%%%%%%%%%%%%%%%%%
%
\subsection{Spall criterion and spall strength}
\label{sec:SpallCriterion}
The \emph{spall strength} is the quantity of primary interest in
dynamic ductile damage experiments. It is defined as the maximum
\emph{macroscopic} stress $\overline{\sigma}_{\rm s}$ sustained by
the material during the damage process. Given the relationship
$\overline{\sigma}_{\rm m}=\overline{\sigma}_{\rm m}(\sigma_{\rm
m})$
%%
%% PAGE 10
%%
between the macroscopic stress $\overline{\sigma}_{\rm m}$
and the microscopic stress $\sigma_{\rm m}$ in non-perturbed,
uniformly loaded, regions of the matrix (see Section
\ref{subsec:SimplifiedGrowthModel}), the macroscopic spall
strength $\overline{\sigma}_{\rm s}$ can be obtained as
$\overline{\sigma}_{\rm s} \equiv \overline{\sigma}_{\rm
m}(\sigma_{\rm s})$, where $\sigma_{\rm s}$ is the microscopic
spall stress solution of
\begin{eqnarray}
    \label{eqn:SpallCriterion}
    \frac{\mathrm{d}\overline{\sigma}_{\rm m}}
    {\mathrm{d}\sigma_{\rm m}}(\sigma_{\rm m}=\sigma_{\rm s}) = 0.
\end{eqnarray}
The spall strength $\overline{\sigma}_{\rm s}$ corresponds to the
maximum stress in the plot of Fig.~\ref{fig:Figure8}.
For the ramp load solution,
Eqs.~(\ref{eqn:ProbabilityOfObscurationClosedForm}) and
(\ref{eq:constit}), the derivative in (\ref{eqn:SpallCriterion})
is carried out using $t=\sigma_{\rm m}/\dot{\sigma}$ and
$\delta=1$ or $\beta$ in the solution. It vanishes for a
dimensionless critical time for spall
\begin{eqnarray}
   \label{eq:ts}
    \widetilde{t}_s = \left\{ [m+3(\alpha+1)] B\bigl(m,3(\alpha+1)\bigr)
    \right\}^{-1/[m+3(\alpha+1)]}
\end{eqnarray}
and a corresponding macroscopic spall strength
\begin{eqnarray}
   \label{eq:ss}
    \overline{\sigma}_s = \sigma_c \left\{
    [m+3(\alpha+1)] B\bigl(m,3(\alpha+1)\bigr) e
    \right\}^{-1/[m+3(\alpha+1)]},
\end{eqnarray}
where $e = \exp(1)$, $\widetilde{t}_s = t/t_{\rm cc}$,
$\sigma_c=\sigma_{\rm cc}$ for the upper ``bound'', and
$\widetilde{t}_s = t/t_{\rm ci}$, $\sigma_c=\sigma_{\rm ci}$  for the
lower ``bound''. At the spall point, the damage parameter is equal
\emph{for both ``bounds''} to
\begin{eqnarray}
    \overline{D}_{s} = 1 - \exp\{-1/[m+3(\alpha+1)]\}.
\end{eqnarray}

%%%%%%%%%%%%%%%%%%%%%%%%%%%%%%%%%%%%%%%%%%%%%%%%%%%%%%%%%%%%%%%%%%%%%%%%%%%%%%%%%%%%%
%%%%%%%%%%%%%%%%%%%%%%%%%%%%%%%%%%%%%%%%%%%%%%%%%%%%%%%%%%%%%%%%%%%%%%%%%%%%%%%%%%%%%
%
\section{Analyses of experiments on tantalum}
\label{sec:TantalumExperiments}
%
%%%%%%%%%%%%%%%%%%%%%%%%%%%%%%%%%%%%%%%%%%%%%%%%%%%%%%%%%%%%%%%%%%%%%%%%%%%%%%%%%%%%%
%
\subsection{The material}
\label{subsec:TheMaterial}
Tantalum is a transition metal of great interest for studying
dynamic ductile damage mainly because of its high mass density
(16,660 kg/m$^3$), good dynamic strength and very high ductility
in wide strain rate and temperature ranges. The samples used
herein are machined from $5$ mm thick cross-rolled and fully
recrystallized plates. Advanced elaboration process and heat
treatment resulted in a very high purity material (99.98 wt\%).
The main (embrittling) impurities are 15 wt ppm O, 15 wt ppm C and
less than 10 ppm N, with a homogeneous microstructure
characterized by equiaxed grains of typical size $90 \mu$m, and a
weak residual texture. Either optical microscopy,
SEM or SIMS examinations did not reveal any localized
heterogeneity down to a $\sim 5$  $\mu$m scale, namely no
second-phase hard particle nor impurity gradient at grain
boundaries. The lack of preferable nucleation sites has been
revealed by dynamic tensile tests on smooth and notched
axisymmetric samples, where failure does occur in any case by
ultimate thinning of the elongated ligament rather than through
inclusion-induced damage, for stress triaxialities ranging from
0.3 to 1 (Roy, 2003). This material is consequently an almost
ideal polycrystal for studying \emph{homogeneous} ductile
nucleation (Roy, 2003).

Mechanical properties of tantalum have been carefully determined
from ultrasonic measurements, quasi-static and dynamic uniaxial
testing on both as received and shocked material (Roy, 2003).
During the release stage following the initial shock compression,
tantalum behaves roughly as an isotropic elastic perfectly plastic
medium (Juanicotena, 1998; Roy, 2003). This holds both at the
macroscopic scale during release wave interaction when no damage
occurs and at the mesoscopic scale around growing voids, where high
strain-rate gradients are roughly balanced by thermal softening at
large strain. The relevant properties of tantalum in the range of
stress and strain states of interest are summarized in Table
\ref{table:Table1}.
\begin{table}[thb]
    \centering
    \includegraphics[width=18cm]{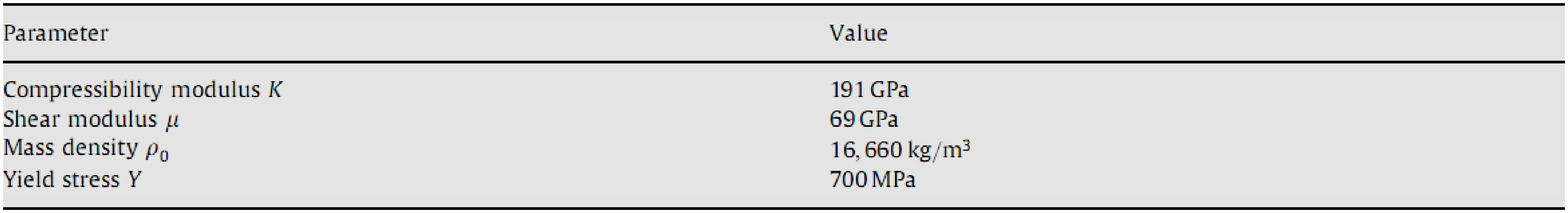}
    \caption{Tantalum material parameters.}
    \label{table:Table1}
\end{table}
%\begin{table}[thb]
%    %\vskip1.0cm
%\begin{center}
%        \begin{tabular}{|l|c|}
%            \hline
%            \hline
%            Parameter & Value \\
%            \hline
%            Compressibility modulus $K$ & 191 GPa \\
%            Shear modulus $\mu$         & 69 GPa \\
%            Mass density  $\rho_0$      & 16,660 kg/m$^3$ \\
%            Yield stress $Y$            & 700 MPa \\
%            \hline
%            \hline
%        \end{tabular}
%\end{center}
%        \caption{Tantalum material parameters.}
%        \label{table:Table1}
%\end{table}
%

Twenty-two plate impact experiments (Nicollet \coll, 2001; Roy, 2003; Llorca and Roy, 2003; Bontaz-Carion and Pellegrini, 2006) were
performed and/or analyzed for the present paper. Impact velocity,
flyer plate material and flyer plate thickness were selected as
relevant parameters for varying both shock pressure and pulse
duration, and are summarized in Table 2. This
essentially induces variations in the position of the plane of
maximum tensile stress (the spall plane) and in the mean and
maximum achievable tensile stress state along this plane. The
diagnostics used to study the condition for damage and spall to
occur are Doppler laser interferometry to record the velocity of
the target free surface (overall structural response of the sample
plate) and qualitative and quantitative metallurgical analyses of
the soft-recovered samples. The most significant results derived
from this microstructural examination have been reported elsewhere
(Roy, 2003; Llorca and Roy, 2003; Nicollet \coll, 2001; Bontaz-Carion and Pellegrini, 2006).
%%
%% PAGE 11
%%
%
\begin{figure}[ht]
    \centering
    \includegraphics[width=8.5cm]{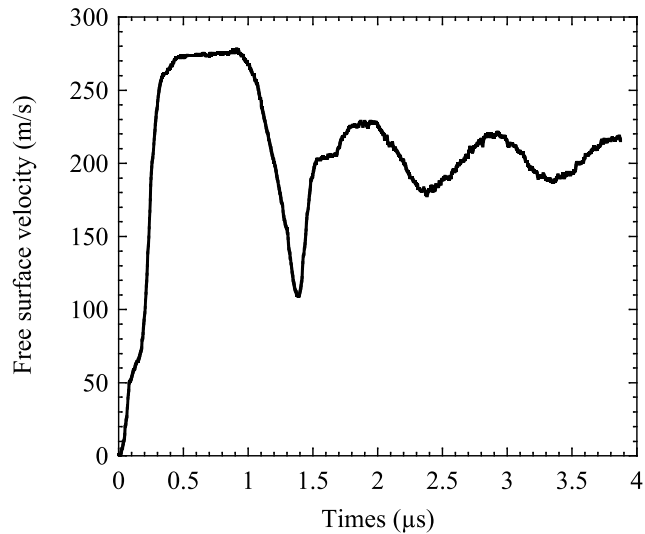}
    \caption{Example of a free-surface velocity record exhibiting a pull-back signal (Roy, 2003).}
    \label{fig:Figure9}
\end{figure}
%
%%%%%%%%%%%%%%%%%%%%%%%%%%%%%%%%%%%%%%%%%%%%%%%%%%%%%%%%%%%%%%%%%%%%%%%%%%%%%%%%%%%%%
%
\subsection{Data extraction}
\label{subsec:Identification}
Time-resolved in situ measurements in the spall plane are not yet
possible, and an inverse methodology must be adopted. As a result,
data extraction is performed from free surface velocity records.
In the spall plane, progressive damage induces local relaxation
waves whose macroscopic consequence is a so-called \emph{pullback
signal} (see Fig.~\ref{fig:Figure9}). For first-order estimations
of relevant data (spall plane location, spall strength, critical
time to fracture), a simple analytical elastic method is often
used (Romanchenko and Stepanov, 1980). This method proves
successful at low shock pressure (lower than the material dynamic
yield strength) or at high pressure (when elastic behavior can be
neglected regarding plastic hydrodynamic component) (Meyers,
1994). This is definitely not the case for tantalum, whose dynamic
yield strength (or
%%
%% PAGE 11
%%
\emph{Hugoniot Elastic Limit}) is known to be
less than an order of magnitude lower than its spall strength in
the range of loading paths of interest.

Accurate data extraction requires an analysis of the complex wave
pattern induced by the plate impact. One- and two-dimensional
numerical simulations have consequently been performed using the
Lagrangian explicit hydrocode Hesione (a proprietary code of the
Commissariat \`a l'\'Energie Atomique). In order to extract the
thermomechanical fields in the region of interest (the spall
plane) as accurately as possible, a tabulated equation of state
and a Preston {\coll} (2003) viscoplastic constitutive law are used (Juanicotena, 1998).

Within the relatively low shock pressure range investigated here
(low temperature increase and weak plastic strain during shock and
release at the macroscopic scale), these relationships do predict an
essentially elastic perfectly plastic behavior \emph{during
unloading}, consistent with the analytical parameters summarized
in Table \ref{table:Table1}. These relationships fitted from
dedicated experimental databases on shock and uniaxial compression
behavior of this tantalum grade (Roy, 2003), yield very good
correlation with experimental results used in this study regarding
shock and release behavior. A fracture criterion is added, leading
to instantaneous mesh opening at a given tensile stress threshold
(spall strength), and fitted numerically for each simulated
experiment.

This numerical procedure is sufficient to extract
the following data from free-surface velocity records: the stress rate in the matrix, the critical time and the spall strength. In order for
the extraction procedure to be as accurate as possible, two
features are particularly sought in matching numerical results and
free-surface velocity records, namely the minimum velocity
preceding pullback signal\footnote{We emphasize that a completely
fractured plane at the macroscopic scale is not a necessary
condition for pullback-type free surface velocity evolution, as
highlighted by Llorca and Roy (2003) and Roy (2003). Primary
internal energy release leading to pullback velocity (early
re-acceleration) has been experimentally shown to be initiated in
the vicinity of the spall plane at a given low incipient damage
level. This is consistent with the basic hypothesis of the spall
criterion developed in part 3.} and the subsequent ringing
velocity frequency, which suggests efficient prediction of both
spall plane position (which was compared with the experimental
value for some experiments), effective maximum tensile stress and
associated critical time, in a far more accurate way than using
the simplified analytical method presented by Roy (2003).
\begin{table}[thb]
    \centering
    \includegraphics[width=18cm]{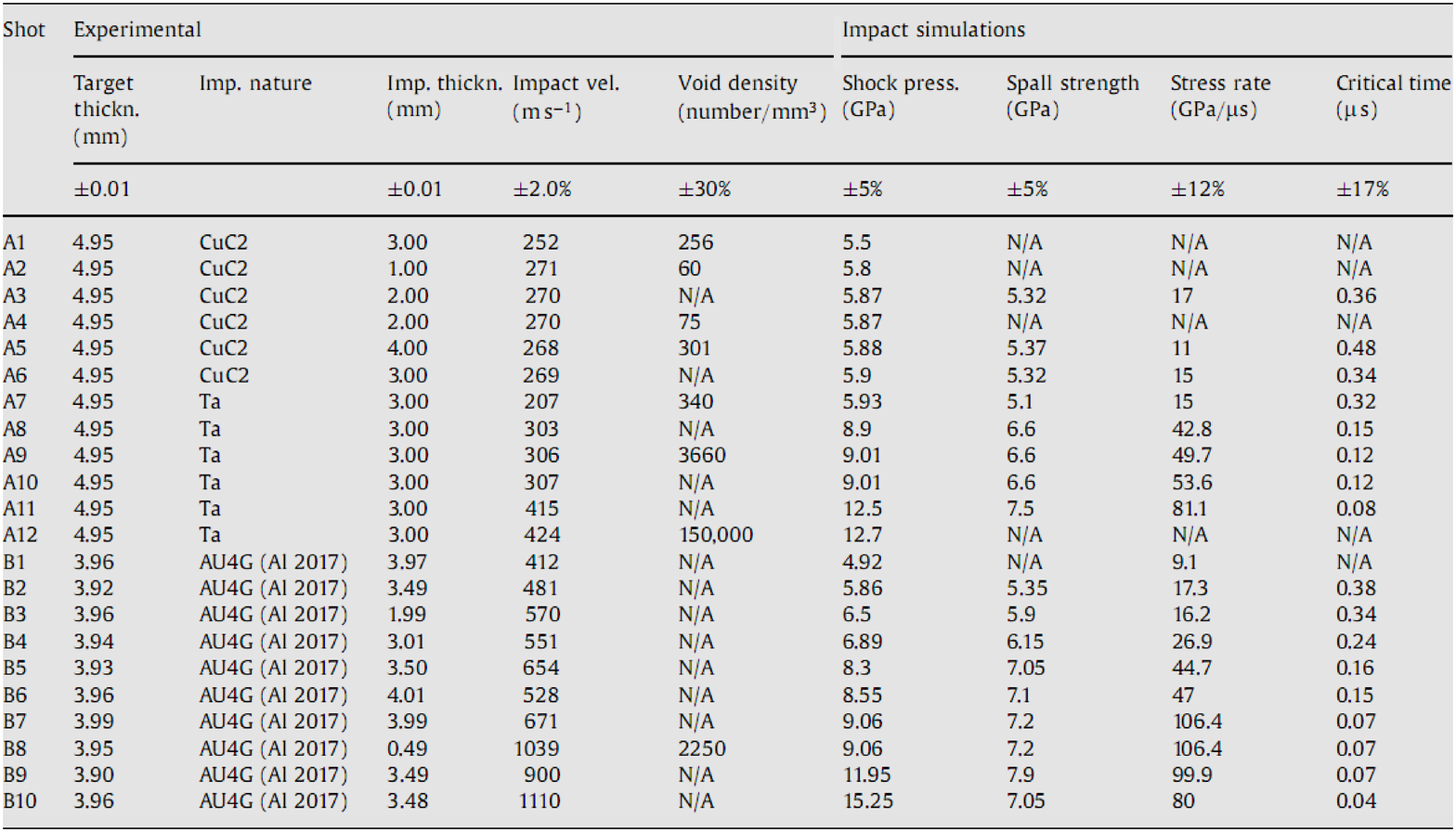}
    \caption{Parameters of shock experiments. Shots A1--A12 (resp.~B1--B10) are those of Roy (2003) and Llorca and Roy (2003) (resp.~Nicollet \coll, 2001; Bontaz-Carion and Pellegrini, 2006). Third row: standard uncertainties. Sixth column: voids densities measured by image analysis on recovered samples. N/A indicates unavailable data.}
    \label{table:Table2}
\end{table}
In this fashion, critical time (spall criterion activation) and
mean tensile stress rate are derived from numerical stress history
prediction at the spall plane before fracture. The corresponding
values are summarized in Table~\ref{table:Table2}. Associated
error estimations are derived from numerical investigation of
impact velocity uncertainty, mesh size, and artificial viscosity
sensitivity of the calculated spall strength for some typical
experiments. For some of these experiments, quantitative relevant
damage activation measures were also derived using metallurgical
observation of sample slices coupled with optical profilometry and
image analysis (Roy, 2003) for an estimation of the three-dimensional damage state. In particular, the volume density of nucleated voids could be measured in the vicinity of the spall plane. These values are also given in Table \ref{table:Table2}.
%
%%%%%%%%%%%%%%%%%%%%%%%%%%%%%%%%%%%%%%%%%%%%%%%%%%%%%%%%%%%%%%%%%%%%%%%%%%%%%%%%%%%%%
%
\subsection{Identification and validation}
\label{subsec:Validation}
\begin{figure}[ht]
    \centering
    \includegraphics[width=8.5cm]{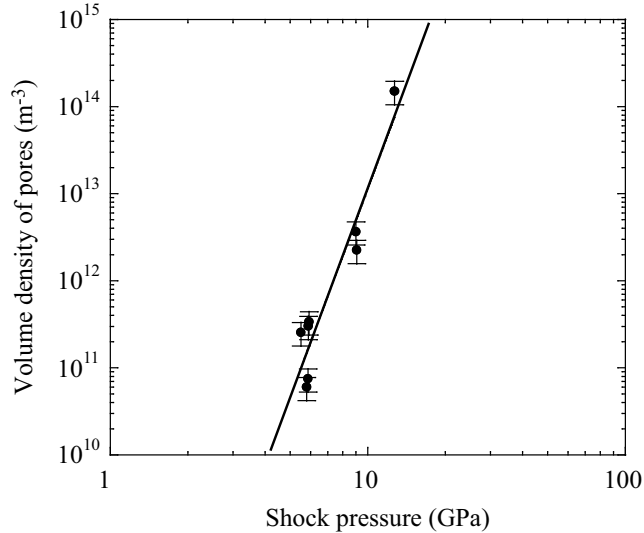}
    \caption{Volume density of pores $n_{\rm tot}$
    vs.~shock pressure for tantalum.
    The solid symbols are experimental points and the line is the
    best fit of Eq.~(\ref{eqn:StressDependentDensity}).}
    \label{fig:Figure10}
\end{figure}

Fig. \ref{fig:Figure10} shows the change of volume density of
nucleated voids $n_{\rm tot}$, Eq.~(\ref{eqn:StressDependentDensity}), as a function of the
shock pressure. This plot is restricted to data obtained from
shots A1--A5, A7, A9, A12 and B8 only. These shots involve only
moderate pressures so that void coalescence presumably remains
limited. Moreover, in the impact configurations considered, the
shock pressure is equal to the negative of $\sigma_{\rm m}$, the
maximum stress in the matrix, which takes place in non-inhibited
regions that exist whenever coalescence is marginal. The
assumption of a constant stress rate pulse (i.e., ramp load) is
applied to tantalum to determine the Weibull parameters of
Eq.~(\ref{eqn:StressDependentDensity}). The best power-law fit
displayed in Fig.~\ref{fig:Figure10} provides an exponent $m = 8$,
a moderate value indicative of weak scatter in nucleation levels.
A value of $\sigma_{0} = 700$ MPa for the scaling stress equal to
elastic limit is used, whence the density $n_0 = 7.9 \times
10^{-6}$ mm$^{-3}$ is obtained.
\begin{figure}[ht]
    \centering
    \includegraphics[width=8.5cm]{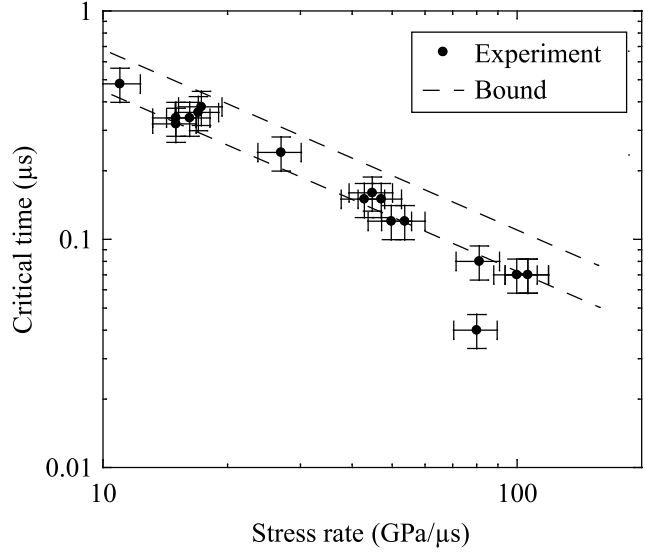}
    \caption{Critical time vs.~stress rate. The solid symbols are
    experimental points and the dashed lines are the ``bounds''
    built from Eq.~(\ref{eq:ts}) with $\alpha=1/2$, and $m =8$
    determined from Fig.~\ref{fig:Figure10}.}
    \label{fig:Figure11}
\end{figure}
\begin{figure}[ht]
    \centering
    \includegraphics[width=8.5cm]{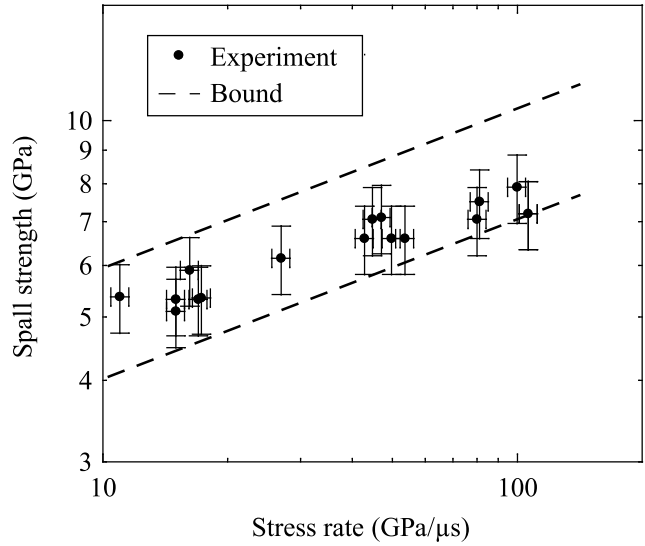}
    \caption{Spall strength vs.~stress rate for tantalum. The solid symbols are experimental points and the dashed lines are the bounds
    built from Eqs.~(\ref{eq:ts}) and (\ref{eq:ss}) with $\alpha=1/2$, and $m =8$ determined from Fig.~\ref{fig:Figure10}.}
    \label{fig:Figure12}
\end{figure}
Upper and lower theoretical ``bounds'' for the critical time
vs.~stress rate obtained from (\ref{eq:ts}) with $\alpha=1/2$
and $m = 8$ are displayed in Fig.~\ref{fig:Figure11}. Almost all
experimental points are seen to lie within these ``bounds''.
Besides the overall trend is consistent with the slope of the
latter.
Likewise, upper and lower theoretical bounds derived from
(\ref{eq:ss}) with $\alpha=1/2$ and $m = 8$ are compared to
experimental data in Fig.~\ref{fig:Figure12} in log--log scale,
which illustrates the power-law increase of the spall strength
with the stress rate. Though the experimental points are linearly
correlated with a slope lower than that of the bounds, they lie
between the latter in the considered range of loadings, which is
quite satisfactory. Thus, the rate sensitivity of the spall
strength can be described by the present model with no need to
incorporate a time-dependent constitutive equation of the matrix.
%
%%%%%%%%%%%%%%%%%%%%%%%%%%%%%%%%%%%%%%%%%%%%%%%%%%%%%%%%%%%%%%%%%%%%%%%%%%%%%%%%%%%%%
%%%%%%%%%%%%%%%%%%%%%%%%%%%%%%%%%%%%%%%%%%%%%%%%%%%%%%%%%%%%%%%%%%%%%%%%%%%%%%%%%%%%%
%
\section{Analyses of data on aluminum and magnesium}
\label{sec:AluminumAndMagnesium}
Kanel \coll\ (1996) performed experiments on aluminum and
magnesium. In both cases, the spall strength was shown to be
approximated by a power-law of the strain rate. In the present
analysis, as in all the developments derived herein, the effect of
the temperature is ignored. Consequently, only experiments
performed at ambient are considered. By using $\sigma_{\rm
s}=\dot{\sigma}\, t_{\rm s}=\widetilde{t}_{\rm s}
\dot{\sigma}\,t_c\propto \dot{\sigma}\,t_c$, any of
Eq.~(\ref{eqn:CharacteristicParametersObscurationVolume}) or
(\ref{eqn:CharacteristicParametersCavityVolume}) for $t_c$
vs.~$\dot{\sigma}$, and the proportionality $\dot{\sigma}\propto
\dot{\varepsilon}$ (of elastic origin, and legitimate in
non-relaxed regions of uniform $\sigma_m$), the following
strain-rate dependence is obtained for the microscopic spall
strength
\begin{eqnarray}
    \sigma_{\rm s} \propto \dot{\varepsilon}^{\eta}
\end{eqnarray}
%%
%% PAGE 14
%%
with
\begin{eqnarray}
    \eta = \frac{3}{m+3(\alpha+1)},
\end{eqnarray}
where $\dot{\varepsilon}$ denotes the average strain rate in the
experiments. In this expression, the only unknown is the modulus
$m$, provided a value of $\alpha = 1/2$ is chosen as in the
previous experiments on tantalum. For aluminum, a value $\eta =
0.059$ is found, which would lead to a value of $m = 46$ and for
magnesium, $\eta = 0.072$ so that $m = 37$. These two (high)
values of $m$ are an indication of a small scatter  in terms of
nucleation level when compared to tantalum (Table
\ref{table:WeibullParameters}) for which a gradual and
scattered nucleation was observed.
\begin{table}[thb]
%\vskip1.0cm
%    \begin{center}
%        \begin{tabular}{|c|c|c|c|}
%            \hline
%            \hline
%            Parameter & Tantalum & Aluminum & Magnesium \\
%            \hline
%            $\alpha$ & 0.5 & 0.5 & 0.5 \\
%            Weibull modulus m & 8 & 46 & 37 \\
%            \hline
%            \hline
%        \end{tabular}
%  \end{center}
    \centering
    \includegraphics[width=18cm]{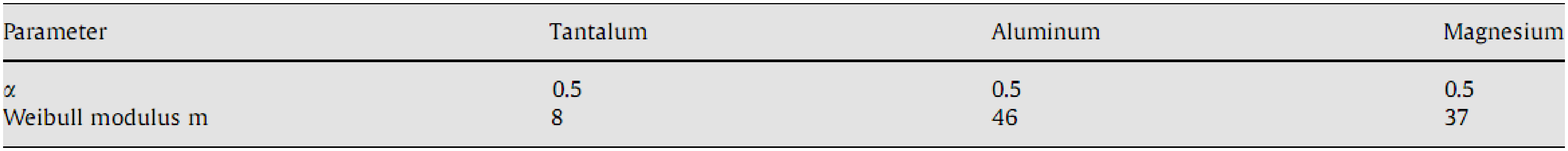}
     \caption{Nucleation parameters for tantalum, aluminum and magnesium.}
     \label{table:WeibullParameters}
\end{table}
%
%%%%%%%%%%%%%%%%%%%%%%%%%%%%%%%%%%%%%%%%%%%%%%%%%%%%%%%%%%%%%%%%%%%%%%
%
\section{Conclusion and perspectives} \label{Conclusion}
We proposed a probabilistic model for nucleation and growth in
ductile fracture, using Poisson--Weibull statistical concepts,
which are usually applied to brittle materials. We showed, through
analyses of several sets of experimental data in spall
experiments, that these concepts are well suited to describing
ductile fracture as well. In particular, we arrived at a simple
explanation for the power-law dependence of the spall strength
vs.~strain rate observed by Kanel \coll\ (1996). The proposed model
makes use of a velocity-dependent extension of the concept of
cavitation stress in metals. Though it has been presented, for
simplicity, in the framework of ideal plasticity, expressions of
cavitation thresholds that account for hardening are available
(Mandel, 1966; Bishop and Hill, 1945), and could be easily
appealed to. Investigations of the influence of hardening on the
present nucleation theory, as well as that of viscoplastic
behavior, are left to future work. Also, next steps should consist
in implementing the full model in a finite-element hydrocode, and
in extending its range of validity to the coalescence regime.

\vspace{0.8cm}

%
%%%%%%%%%%%%%%%%%%%%%%%%%%%%%%%%%%%%%%%%%%%%%%%%%%%%%%%%%%%%%%%%%%%%%%%%%%%%%%%%%%%%%
%%%%%%%%%%%%%%%%%%%%%%%%%%%%%%%%%%%%%%%%%%%%%%%%%%%%%%%%%%%%%%%%%%%%%%%%%%%%%%%%%%%%%
%
{\bf Acknowledgments}
\vspace{0.5cm}

This work was partially funded by the CNRS. The authors wish to
thank Prof.\ An\-dr\'{e} Dragon for stimulating discussions on
spallation, and Dr.\ Jo\"{e}l\-le Bontaz-Carion for having made
available to us the experimental data of Nicollet \coll\ (2001).
%
%%%%%%%%%%%%%%%%%%%%%%%%%%%%%%%%%%%%%%%%%%%%%%%%%%%%%%%%%%%%%%%%%%%%%%%%%%%%%%%%%%%%%
%%%%%%%%%%%%%%%%%%%%%%%%%%%%%%%%%%%%%%%%%%%%%%%%%%%%%%%%%%%%%%%%%%%%%%%%%%%%%%%%%%%%%
\bigskip

{\bf\large Appendix}

\begin{appendix}
\numberwithin{equation}{section}
%%%%%%%%%%%%%%%%%%%%%%%%%%%%%%%%%%%%%%%%%%%%%%%%%%%%%%%%%%%%%%%%%%%%%%%%%%%%%%%%%%%%%
%%%%%%%%%%%%%%%%%%%%%%%%%%%%%%%%%%%%%%%%%%%%%%%%%%%%%%%%%%%%%%%%%%%%%%%%%%%%%%%%%%%%%
\section{Weibull distribution}
\label{appendix:Weibull}
The probability of finding {\em at least} one nucleation site (i.e.,
the ``weakest link") in a uniformly loaded domain $\Omega$ is
\begin{eqnarray}
    \label{eqn:Poisson-Weibull}
    P(N\geq 1,\Omega)=1-P(N=0,\Omega)=1-e^{-V\, n_0\left(\langle\sigma_{\rm m}\rangle/\sigma_{0}\right)^{m}}.
\end{eqnarray}
When the domain is not uniformly loaded, we instead have
\begin{eqnarray}
    \label{eqn:Weibull}
    P(N \geq 1,\Omega) = 1-e^{-V_\mathrm{eff}\,n_0\left(\langle
    \sigma_{M}\rangle/\sigma_{0}\right)^{m}},
\end{eqnarray}
where $V_{\mathrm{eff}}$ denotes the effective volume (Davies,
1973)
\begin{eqnarray}
   \label{eqn:Z_eff}
   V_\mathrm{eff} = \int_{\Omega}
   \mathrm{d}^3\!x\,\left[\frac{\sigma_{\rm m}(\mathbf{x})}{\sigma_{\rm M}}\right]^{m}
   \quad\mbox{with}\quad\sigma_{\rm M} = \max_{\Omega}
   \sigma_{\rm m}(\mathbf{x}).
\end{eqnarray}
Eqs.\ (\ref{eqn:Poisson-Weibull}) and (\ref{eqn:Weibull}) are
the Weibull model (Weibull, 1951) written in the context of
ductile damage (see also Czarnota \coll, 2006).
%
%%%%%%%%%%%%%%%%%%%%%%%%%%%%%%%%%%%%%%%%%%%%%%%%%%%%%%%%%%%%%%%%%%%%%%%%%%%%%%%%%%%%%
%%%%%%%%%%%%%%%%%%%%%%%%%%%%%%%%%%%%%%%%%%%%%%%%%%%%%%%%%%%%%%%%%%%%%%%%%%%%%%%%%%%%%
%
\section{Derivation of
Eqs.~(\ref{eqn:CdotVersusAdot}) and
(\ref{eqn:AdotVersusPressure})} \label{appendix:ForrestalLuk}
%
%%%%%%%%%%%%%%%%%%%%%%%%%%%%%%%%%%%%%%%%%%%%%%%%%%%%%%%%%%%%%%%%%%%%%%%%%%%%%%%%%%%%%
%
\subsection{Preliminaries}
We detail here the steps leading to
Eqs.~(\ref{eqn:CdotVersusAdot}) and
(\ref{eqn:AdotVersusPressure}), in the stationary growth regime
studied by Forrestal and Luk (1988). In this one-dimensional
spherical approach, a cavity of radius $a(t)$ grows at constant
velocity $\dot{a}$ in an infinite elasto-plastic medium submitted
to an initially uniform hydrostatic stress state $\sigma_{\rm
m}(t)$. This growth perturbs the stress field within a partially
relaxed volume of radius $r=b(t)\equiv c_L\,t$, where $c_L =
\sqrt{(K +4\mu/3)/\rho_0}$ is the velocity of longitudinal elastic
waves. The front $b(t)$ separates the outer medium at rest in a
state of uniform stress, from the inner perturbed region expanding
with the growing void. The inner region is divided into an
external elastic shell $c(t) \leq r \leq b(t)$, and a shell at
yield that surrounds the cavity, $a(t) \leq r \leq c(t)$.

In the steady-state growth regime where $c(t)=\dot{c}\,t$, a
self-similar solution for the radial displacement $u$ is sought
for in the form $u(r,t)=c(t)\,\widetilde{u}(\xi)$. There,
$\xi(r,t)=r/c(t)$ is the scaled radial coordinate, and
$\widetilde{u}(\xi)$ is the scaled displacement. Moderate stress
is assumed so as to neglect (i) density variations in the elastic
shell, (ii) non-linear elasticity, and (iii) convection terms
(Forrestal and Luk, 1988). Using $\dot{\xi} = - \xi\,
\dot{c}/c$, the velocity and acceleration read
\begin{eqnarray}
   \label{eq:vit}
    \dot{u} &=& \left(\widetilde{u}- \xi\, \widetilde{u}\,'\right) \dot{c}, \\
    \ddot{u}&=& \xi^2 \widetilde{u}\,''\,
    \dot{c}\,{}^2/c.
\end{eqnarray}
Eq.\ (\ref{eq:vit}) provides the scaled velocity
$\widetilde{v}(\xi) \equiv \dot{u}(r,t)/\dot{c}$.

For further use, we introduce the scaled yield stress, shear
modulus, applied hydrostatic stress, and density, respectively, as
$y\equiv Y/ K$, $g\equiv 2\mu/K$, $\widetilde{\sigma}(t)\equiv
\sigma_{\rm m}(t)/ K$, $\widetilde{\rho}(\xi)\equiv
\rho(r,t)/\rho_0$, where $\rho_0$ is the reference material
density. In usual metals,
\begin{eqnarray}
\label{eq:approx} y\ll g \lesssim 1.
\end{eqnarray}
%
%%%%%%%%%%%%%%%%%%%%%%%%%%%%%%%%%%%%%%%%%%%%%%%%%%%%%%%%%%%%%%%%%%%%%%%%%%%%%%%%%%%%%
%
\subsection{Elastic shell}
Combining linear elasticity relationships and the momentum equation
\begin{equation*}
\partial_r \sigma_r+(2 / r)(\sigma_r - \sigma_\theta)=
\rho \ddot{u},
\end{equation*}
where $\sigma_r$ and $\sigma_\theta$ are, respectively, the radial
and hoop stresses, and introducing $\gamma_L\equiv\dot{c}/c_L$,
yields the differential equation
\begin{eqnarray}
    \label{eqn:IntegratedElasticSolution}
    \left(1-\gamma_L^2\,\xi^2\right) \widetilde{u}\,'' + \left(2/\xi^2\right) \left(\xi\,\widetilde{u}\,'- \widetilde{u}\,\right) =
    0.
\end{eqnarray}
Its solution is of the form $
\widetilde{u}(\xi)=C^{(1)}\xi+C^{(2)}(1-3\gamma_L^2\,\xi^2)/\xi^2$,
where the integration constants $C^{(1,2)}$ are found from
boundary conditions. The first one is
$\widetilde{u}(\xi=1/\gamma_L)=\sigma_{\rm m}/(3K\gamma_L)$, and
stems from the applied external boundary traction. The second one
is $\widetilde{u}(\xi) /\xi - \widetilde{u}\,'(\xi) |_{\xi=1} = Y
/ (2\mu)$, which expresses the yield condition $\sigma_\theta -
\sigma_r = Y$ (tensile case) at the elastic--plastic boundary. The
solution for $\xi \in (1,1/ \gamma_L)$ is then
\begin{eqnarray}
    \label{eqn:ElasticDisplacement}
    \widetilde{u}^e= \frac{\widetilde{\sigma}}{3} \xi + \frac{Y}{6\mu} \frac{(1-\gamma_L\, \xi)^2(1 + 2 \gamma_L\,\xi)}{(1 -
    \gamma_L^2)\,
    \xi^2}.
\end{eqnarray}
Denoting by $\nu$ the Poisson ratio, the corresponding radial
stress reads
\begin{eqnarray}
    \label{eqn:ElasticRadialStress}
    \widetilde{\sigma}_r^e = \widetilde{\sigma} - \frac{2 y}{3} \frac{(1 - \gamma_L\,\xi)\left[(1 - 2 \nu)(1 + \gamma_L\, \xi)
        + (1 + \nu) \gamma_L^2\, \xi^2\right]}{(1 - 2 \nu)\left(1 - \gamma_L^2\right) \xi^3}.
\end{eqnarray}
%
%%%%%%%%%%%%%%%%%%%%%%%%%%%%%%%%%%%%%%%%%%%%%%%%%%%%%%%%%%%%%%%%%%%%%%%%%%%%%%%%%%%%%
%
\subsection{Plastic shell}
Mass conservation, namely, $\partial_t \rho+ [\partial_r + (2 / r)]
(\dot{u}\,\rho) = 0$, provides
\begin{eqnarray}
    \label{eqn:MassConservation}
    \widetilde{v}\,'+ (2/\xi) \widetilde{v}= \left( \xi-\widetilde{v}\right)
    \widetilde{\rho}\,'/\widetilde{\rho}.
\end{eqnarray}
Introduce now the plastic velocity $c_P = \sqrt{K / \rho_0}$, and
(after $\gamma_L$) another scaling of $\dot{c}$ as $\gamma_P
\equiv \dot{c}/c_P$. The yield condition $\sigma_\theta - \sigma_r
= Y$, combined with linear elasticity in the form
$\mathop{\mathrm{Tr}}\sigma$ $=$ $\sigma_r + 2 \sigma_\theta$ $=$
$3 K$ $\left(\rho_0/\rho-1\right)$, gives
\begin{eqnarray}
    \label{eqn:RadialStressVersusDensity}
    \partial_r \sigma_r= - K \left(\rho_0/\rho^2\right) \partial_r \rho.
\end{eqnarray}
Using (\ref{eqn:RadialStressVersusDensity}) in the momentum
equation then provides
\begin{eqnarray}
    \label{eqn:Density}
    \xi\,\widetilde{\rho}\,'=\left(\gamma_P^2\,\xi^2\,\widetilde{v}\,'\,\widetilde{\rho}-2 y\right)
    \widetilde{\rho}^2.
\end{eqnarray}
Eliminating $\widetilde{\rho}'(\xi)$ from
Eqs.~(\ref{eqn:MassConservation}) and (\ref{eqn:Density}) then
yields
\begin{eqnarray}
    \label{eqn:MasterVelocityEquation}
        \left[1 - \xi\,\gamma_P^2\,\widetilde{\rho}\,^2\,\left(\xi - \widetilde{v}\right) \right] \widetilde{v}\,'
            + 2\left(1 - y\,\widetilde{\rho}\right)
            (\widetilde{v}/\xi)
             = - 2y\, \widetilde{\rho}.
\end{eqnarray}
Eqs.\~(\ref{eqn:Density}) and
(\ref{eqn:MasterVelocityEquation}) constitute a system for
$\widetilde{\rho}$ and $\widetilde{v}$, should variations in
$\rho$ be accounted for. Upon neglecting their higher order
influence in $\widetilde{v}$ at moderate stress, and assuming
$y\ll 1$, see (\ref{eq:approx}),
Eq.~(\ref{eqn:MasterVelocityEquation}) reduces to
\begin{eqnarray}
    \label{eqn:VelocityEquation}
    \left(1 - \gamma_P^2\, \xi^2\right) \widetilde{v}\,'+
    2(\widetilde{v}/\xi) = -2y.
\end{eqnarray}
Note that this equation also assumes that $\widetilde{v}(\xi) \ll
\xi$, which is satisfied if the material velocity is much lower
than the velocity of the void boundary. However, finite-element
calculations of void expansion (Roy, 2003) indicate that this
assumption is expected to hold everywhere except near the void
boundary where the velocity gradient is highest. The difference
induced by neglecting this term on the overall behavior is small
anyway (Forrestal and Luk, 1988; Roy, 2003), see
Fig.~\ref{fig:Figure13} below.

Continuity of the material velocity at the elastic--plastic
interface provides the boun\-da\-ry condition $ \widetilde{v}(1)=
y/g$. Then, the solution of Eq.~(\ref{eqn:VelocityEquation}) in
the interval $\xi$ $\in$ $(a/c,1)$ is
\begin{eqnarray}
\label{eqn:PlasticVelocity}
        \widetilde{v}^p(\xi) = \frac{y}{\gamma_P^2\,\xi^2} \left[ \frac{1 - \gamma_P^2\,\xi^2}{1 - \gamma_P^2}(1+\gamma_P^2/g) - \xi \right]
            + \frac{y}{2 \gamma_P^3\,\xi^2} (1 - \gamma_P^2\,\xi^2)\log\frac{(1 + \gamma_P\,\xi)(1 - \gamma_P)}{(1 - \gamma_P \,\xi)(1 + \gamma_P)}.
\end{eqnarray}
With $\alpha_P \equiv \dot{a}/c_P$ the scaled void growth
velocity, the radial stress in the same interval reads, upon
integrating (\ref{eqn:RadialStressVersusDensity}) and using
(\ref{eqn:MassConservation}) under the above approximations
\begin{eqnarray}
    \label{eqn:PlasticRadialStress}
        \widetilde{\sigma}_r^p (\xi) &=&2 y
        \frac{( \gamma_P\,\xi-\alpha_P)(1+\gamma_P^2/g)}{(1-\gamma_P^2)\alpha_P\,\xi}
        +y \left[\log\frac{\gamma_P^2\,\xi^2(1-\alpha_P^2)}{\alpha_P^2(1-\gamma_P^2\,\xi^2)}
        -\frac{1}{\gamma_P\,\xi}
        \log\frac{(1+\gamma_P\,\xi)(1-\gamma_P)}{(1-\gamma_P\,\xi)(1+\gamma_P)}\right.\nonumber\\
        &&\hspace{+21em}{}+ \left. \frac{1}{\alpha_P} \log\frac{(1+\alpha_P)(1-\gamma_P)}{(1-\alpha_P)(1+\gamma_P)}\right].
\end{eqnarray}
%
%%%%%%%%%%%%%%%%%%%%%%%%%%%%%%%%%%%%%%%%%%%%%%%%%%%%%%%%%%%%%%%%%%%%%%%%%%%%%%%%%%%%%
%
\subsection{Complete and approximate solutions}
\begin{figure}[ht]
    \centering
    \includegraphics[width=10cm]{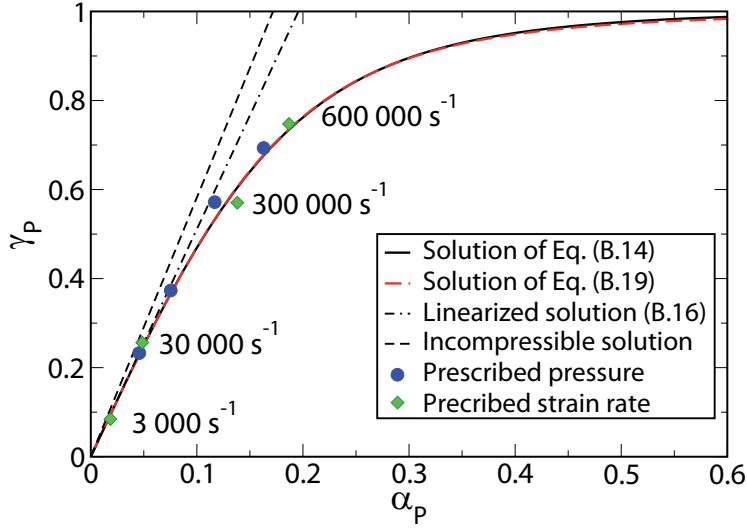}
    \caption{Dimensionless velocity $\gamma_P=\dot{c}/c_P$
    of the plastic zone vs.~dimensionless growth velocity $\alpha_P=\dot{a}/c_P$ of a
    cavity in a compressible and incompressible medium (the value of $c_P$ at finite compressibility is used for all curves).
    Material parameters of tantalum
    (Table \ref{table:Table1}), except for the bulk modulus in the incompressible case.}
    \label{fig:Figure13}
\end{figure}
The equations for the void growth velocity then consist in the
relations $\alpha_P = \gamma_P\, \widetilde{v}^p(\xi=a/c)$ and
$\widetilde{\sigma}_r^e(1)=\widetilde{\sigma}_r^p(1)$. The first
one reads
\begin{eqnarray}
    \label{eqn:CdotVersusAdotExact}
       \alpha_P\frac{\alpha_P^2 / y + 1}{1 - \alpha_P^2} =
       \gamma_P \frac{\gamma_P^2 / g + 1}{1 - \gamma_P^2}
        + \frac{1}{2}\log\frac{(1 + \alpha_P)(1 - \gamma_P)}{(1 - \alpha_P)(1 +
        \gamma_P)},
 \end{eqnarray}
whereas, setting $\kappa\equiv c_L/c_P=(1+2g/3)^{1/2}$, the second
one yields
\begin{eqnarray}
    \label{eqn:PressureVersusAdotExact}
    \frac{\widetilde{\sigma}}{y} = &\frac{2}{3} +2\frac{\kappa^2\,\gamma_P^2 / g}{1 + \kappa\, \gamma_P}
            +2\frac{\gamma_P^2 / g + 1}{1 - \gamma_P^2}
            \left( \frac{\gamma_P}{\alpha_P}-1\right)
            + \frac{1}{\alpha_P} \log\frac{(1 + \alpha_P)(1 - \gamma_P)}{(1 - \alpha_P)(1 + \gamma_P)}
            + \log \frac{\gamma_P^2(1 - \alpha_P^2)}{\alpha_P^2(1 - \gamma_P^2)}.
\end{eqnarray}

Seeking low-order expansions of
Eqs.~(\ref{eqn:CdotVersusAdotExact}) and
(\ref{eqn:PressureVersusAdotExact}), $\dot{a}$ is computed as a
function of $\sigma_{\rm m}$ by first looking for a solution of
Eq.~(\ref{eqn:CdotVersusAdotExact}) in the perturbative form $
\gamma_P = \sum_{k \geq 1} A_k \alpha_P^k$, where the unknowns
$A_k$ are determined order-by-order. To leading order in
$\alpha_P$, the solution is
\begin{eqnarray}
    \label{eqn:CdotVersusAdotLinear}
\gamma_P \simeq \beta^{-1/3}\,\alpha_P,\quad
\beta\equiv\frac{y(g+3/2)}{g(y+3/2)}\simeq \frac{Y}{2\mu} +
\frac{2Y}{3K},
\end{eqnarray}
where the approximated value of $\beta$ stems from
(\ref{eq:approx}).
Next, inserting the expansion into
Eq.~(\ref{eqn:PressureVersusAdotExact}), assuming a relationship
$\widetilde{\sigma}(\alpha_P)$ $=$ $\widetilde{\sigma}_c$
$+\sum_{k\geq 1} B_k \alpha_P^k$ where $\widetilde{\sigma}_c$ and
the $B_k$ are unknowns, and again simplifying the coefficients
with (\ref{eq:approx}), yields
\begin{eqnarray}
    \label{eqn:PressureVersusAdotLinear}
        \widetilde{\sigma}=\frac{2 y}{3}(1-\log \beta)
        +\left[2-O\left((y g^2)^{1/3}\right)\right]\alpha_P^2+O\left(\alpha_P^3\right),
\end{eqnarray}
where the orders of the neglected terms are indicated. Hence
$B_2=2$ in the incompressible limit. We do not reproduce its full
expression, quite involved but easily retrieved with a symbolic
calculator.
The first term in the r.h.s.~is the scaled cavitation stress,
$\widetilde{\sigma}_{\rm cav}=\sigma_{\rm cav}/K$, first computed
by Bishop \coll\ (1945),\footnote{However, their expression,
written in terms of $\mu$ and $Y$, is that of the incompressible
case.} and later on by Mandel (1966) for finite
compressibility under the form $\sigma_{\rm
cav}=(2Y/3)\left\{1+\log E/[3(1-\nu)Y] \right\}$, $E$ being
Young's modulus.

Growth occurs only if $\sigma_{\rm m}>\sigma_{\rm cav}$. Hence
from (\ref{eqn:PressureVersusAdotLinear}), for $\sigma_{\rm
m}\gtrsim\sigma_{\rm cav}$, the pore growth velocity behaves as
\begin{eqnarray}
\label{eq:cavitationbehavior} \dot{a} \sim \dot{a}_0 (\sigma_{\rm m}
/ \sigma_{\rm cav} - 1)^{1/2},
\end{eqnarray}
where $\dot{a}_0\equiv [\sigma_{\rm cav}/(B_2\,\rho_0)]^{1/2}$ is
a characteristic pore growth velocity of the material with
$B_2\simeq 2$. Using the full expressions of $\beta$ and $B_2$, we
obtain for Al, Cu and Ta: $\dot{a}_0\simeq 289$, $224$ and $145$
m/s respectively, and $\sigma_{\rm cav}\simeq 0.11$, $0.89$, and
$2.75$ GPa respectively. For comparison purposes, we note that
$c_P\simeq$ $5092$, $3589$, $3386$ m/s for these materials,
respectively, so that $\dot{a}_0$ is lower than $c_P$ by more than
one order of magnitude. Neglecting compressibility provides, with
$B_2=2$, values of $\dot{a}_0$ lower than the above ones by a
relative error of about $5 \times 10^{-3}$. Though it is strictly
valid for a constant applied stress (since
$\dot{a}=\mathrm{const.}$ by hypothesis), Eq.~(\ref{eq:cavitationbehavior}) nonetheless provides the
leading-order behavior for a time-varying stress, which is the
type of loading considered in Sec.~\ref{sec:simplified} where use
is made of this equation. In practice, transient corrections
mainly consist in damped oscillations around this leading
behavior, as was checked by finite-element calculations, and are
neglected in this work.

In the incompressible limit (where $c_L=c_P=\infty$)
Eq.~(\ref{eqn:CdotVersusAdotExact}) reduces to
$\dot{a}=[Y/(2\mu)]^{1/3}\dot{c}$ (this relation is encapsulated
in the equations of Carroll and Holt (1972) in the limit of
vanishing initial porosity). Hence in general, it is expected that
$\dot{a}\ll \dot{c}\leq c_P$. Combined with the above low-velocity
solution, this suggests the following approximation to
(\ref{eqn:CdotVersusAdotExact})
\begin{eqnarray}
\label{eqn:approx} \alpha_P^3=\beta\, \gamma_P^3 / (1-\gamma_P^2),
\end{eqnarray}
which contains in particular the incompressible limit (where
$\gamma_P\to 0$). This approximation, which preserves
(\ref{eqn:PressureVersusAdotLinear}) up to the neglected terms,
and which can be solved analytically for $\gamma_P$, proves useful
to compute numerically the stress in numerical implementations of
the model.

Fig.~\ref{fig:Figure13} compares
Eq.~(\ref{eqn:CdotVersusAdotExact}) with either stress- or
velocity-driven finite-e\-le\-ment numerical results. These data
points are reasonably well reproduced by the solution of
Eq.~(\ref{eqn:CdotVersusAdotExact}), in spite of the underlying
approximations. The solution to Eq.~(\ref{eqn:approx}) is
indistinguishable to the eye from the latter. Also shown are the
linear approximation (\ref{eqn:CdotVersusAdotLinear}) and the
above incompressible (linear) solution.

It should be noted that the incompressible limiting value $B_2=2$
markedly differs from the value $B_2=3/2$ which one easily deduces
from Carroll and Holt's (1972) incompressible calculation in the
limit of zero initial porosity, where convection is accounted for.
Though a detailed study of the influence of convective terms in
the compressible case lies beyond the scope of this paper, this
difference indicates that convection may be important in
accurately determining the coefficient $\dot{a}_0$ in
(\ref{eq:cavitationbehavior}), the difference between the
approaches concerning a numerical coefficient of order one. Taking
$B_2=3/2$ instead of $2$,
Eq.~(\ref{eqn:PressureVersusAdotLinear}) is compatible with the
work of Molinari and Wright (2005) in the limit of stationary
growth of incompressible materials, and close to the result given
by Tonks \coll\ (2001). Thus, the obtained cavitation threshold
and the general form of this law hold in any case, which is a
sufficient conclusion for the present purpose.
%
%%%%%%%%%%%%%%%%%%%%%%%%%%%%%%%%%%%%%%%%%%%%%%%%%%%%%%%%%%%%%%%%%%%%%%%%%%%%%%%%%%%%%
%%%%%%%%%%%%%%%%%%%%%%%%%%%%%%%%%%%%%%%%%%%%%%%%%%%%%%%%%%%%%%%%%%%%%%%%%%%%%%%%%%%%%
%
\section{Inhibition probability}
\label{appendix:ObscurationProbability}
\numberwithin{equation}{section}
To define the probability that a point $\mathbf{x}$ at a time $t$
be relaxed, it is preferable to invert the problem by looking into
the past of the considered site to know if a cavity is able to
inhibit its nucleation (this method, first proposed by Cahn
(1996), was found independently by two of the present authors
(Denoual \coll, 1997; Denoual, 1998)). Two zones are distinguished.
First, a zone in which the nucleated cavities never inhibit the
considered site (see dashed part of Fig.~5(b) when
$\tau<t$). In the second (complementary) zone, any nucleated
cavity will inhibit $\mathbf{x}$. This zone is referred to as the
\emph{horizon} (Cahn, 1996; Denoual \coll, 1997; Denoual, 1998).

The inhibition probability $P_{\rm inh}(t)$ is written as the
product of the elementary probabilities $\Delta
P_{\not\exists}(\tau)$
\begin{eqnarray}
   1-P_{\rm inh}(t) = \prod_{\tau=0}^t \Delta P_{\not\exists} (\tau),
\end{eqnarray}
where $\Delta P_{\not\exists}(\tau)$ is the probability of finding
no new sites during a time increment $\Delta \tau$ in a zone
$V_{\rm inh} (t-\tau)$. It suffices to apply
Eq.~(\ref{eqn:PoissonDistribution}) with $V = V_{\rm
inh}(t-\tau)$ for an intensity $({\rm d} n_{\rm tot}/{\rm d}\tau)
\{\sigma(\tau)\}\Delta \tau$, since it still is a Poisson
process
\begin{eqnarray}
   \Delta P_{\not\exists}(\tau)= \exp \left[ -\frac{\mathrm{d}
   n_{\rm tot}}{\mathrm{d}\tau} \{\sigma(\tau)\} \Delta\tau V_{\rm inh} (t-\tau)
   \right].
\end{eqnarray}
The probability $P_{\rm inh}(t)$ becomes
\begin{eqnarray}
    1 - P_{\rm inh}(t) = \exp
    \left[- \sum_{\tau = 0}^t \frac{\mathrm{d}n_{\rm tot}}{\mathrm{d}\tau}
    \left\{\sigma(\tau)\right\} \Delta \tau V_{\rm inh} (t - \tau)
    \right].
\end{eqnarray}
In the continuous limit $\Delta\tau\to 0$, rewriting the sum as an
integral eventually yields
Eqs.~(\ref{eqn:ProbabilityOfObscuration}) and
(\ref{eqn:AverageObscurationVolume}).

%%%%%%%%%%%%%%%%%%%%%%%%%%%%%%%%%%%%%%%%%%%%%%%%%%%%%%%%%%%%%%%%%%%%%%%%%%%%%%%%%%%%%
\end{appendix}
%%%%%%%%%%%%%%%%%%%%%%%%%%%%%%%%%%%%%%%%%%%%%%%%%%%%%%%%%%%%%%%%%%%%%%%%%%%%%%%%%%%%%
%

%
%%%%%%%%%%%%%%%%%%%%%%%%%%%%%%%%%%%%%%%%%%%%%%%%%%%%%%%%%%%%%%%%%%%%%%%%%%%%%%%%%%%%%
%
\end{document}